\documentclass[12pt]{article}
\usepackage{a4,epsfig}
\begin{document}

%===================> ADD here your LATEX definitions

%================================================ ========================% 

%################################################## titlepage declaration

\begin{titlepage}

\pagenumbering{arabic}
\vspace*{-1.5cm}
\begin{tabular*}{15.cm}{lc@{\extracolsep{\fill}}r}
%{\bf DELPHI Collaboration} & 
%\hspace*{1.3cm} \epsfig{figure=/afs/cern.ch/user/d/delnote/tex/dolphin_bw.eps,width=1.2cm,height=1.2cm}
&
%===================> DELPHI note number       =====> To be filled <=====%
%DELPHI 2016-001 PHYS XXX   
%========================================================================%
\\
& &
%===================> DELPHI note date         =====> To be filled <=====%
17 January  2020
%========================================================================%
\\
&&\\ \hline
\end{tabular*}
\vspace*{1.0cm}
\begin{center}
\Large 
{\bf \boldmath
%===================> DELPHI note title        =====> To be filled <=====%
Physical interpretation  \\
of the anomalous Cherenkov rings   \\
observed with the DELPHI detector
%========================================================================%
}

\vspace*{1.0cm}
\normalsize { 
%===================> DELPHI note author list  =====> To be filled <=====%
   {\bf V. F. Perepelitsa}\\
   {\footnotesize ITEP, Moscow            }\\ 
   
   {\bf T. Ekelof}\\
   {\footnotesize Department of Physics and Astronomy, Uppsala University}\\
   
   {\bf A. Ferrer}\\
   {\footnotesize IFIC, Valencia University}\\

   {\bf B. R. French}\\
   {\footnotesize bernardfrench@bluewin.ch}\\
%========================================================================%
}
%\vspace*{0.4cm}
\end{center}
\vspace{\fill}
\begin{abstract}
\noindent
%===================> DELPHI note abstract     =====> To be filled <=====%
The results of a search with the DELPHI Barrel RICH for anomalous Cherenkov
rings having radii greater than those produced by ultrarelativistic particles
were reported in our previous paper \cite{prashort}. The search was based 
on the data collected by the DELPHI Collaboration at CERN during the 
LEP1 and LEP2 periods. A detailed study of background sources
capable of producing apparently anomalous rings has been done;
it indicated that the background hypothesis has a low probability
($10^{-3}$ or less). 
An additional strong argument against the background hypothesis was provided 
by the observation of a high degree of correlation between anomalous ring 
radii in the liquid and gaseous radiators in the selected events.
The results obtained are interpreted in this paper in terms of  
observation of faster-than-light particles (tachyons). In the framework of 
this interpretation two peaks in the tachyon mass parameter distribution are 
observed, at $(0.29 \pm 0.01)$~GeV/c$^2$
%, $(0.88 \pm 0.10)$~GeV/c$^2$
 and $(4.6 \pm 0.2)$~GeV/c$^2$.

This work has been performed by the authors following the rules for external
access to the DELPHI archived data, as established in
http://delphiwww.cern.ch/delsec/finalrules/FINALrules011203.pdf

%The findings and conclusions expressed in this material are those
%of the authors alone and do not engage in any way the DELPHI Collaboration.
The opinions, findings and conclusions expressed in this material are those
of the authors alone and do not reflect in any way the views of the
DELPHI Collaboration.

%=========================================================================%
\end{abstract}
%\vspace{\fill}
\end{titlepage}

%\pagebreak

%\begin{titlepage}
%\mbox{}
%\end{titlepage}

%\pagebreak

%\setcounter{page}{1}    

%##################################################################### Text

%==================> DELPHI note text          =====> To be filled <======%
\section{Introduction}
\setcounter{equation}{0}
\renewcommand{\theequation}{1.\arabic{equation}}
We present here a physical interpretation of the anomalous Cherenkov rings 
that have resulted from a search for such rings in the events 
of $e^+ e^-$ interactions 
recorded in the DELPHI detector at LEP (CERN) as described in \cite{prashort}
(hereafter denoted by Paper I). The term ``anomalous rings" stands for rings 
having radii greater than those produced by ultrarelativistic particles 
which are termed here ``standard rings". In this article standard rings are
defined as having the ring radius compatible with 667~mrad in the liquid
radiator and 62~mrad in the gaseous radiator, while the anomalous rings are
defined as having the ring radius exceeding 700~mrad in the liquid radiator and
72~mrad in the gaseous radiator.

The results presented in Paper I are based on data consisting of ca~$10^7$ 
non-hadronic events collected at LEP1 and LEP2, corresponding to an integrated 
luminosity of 0.76~fb$^{-1}$. Interpreted in a straightforward way these 
results suggest the existence of faster-than-light particles (tachyons).
Within this interpretation we provide additional 
information in the present paper favouring this interpretation.

Tachyons became topical some 50 years ago after being introduced to particle
physics in papers \cite{bds,fein} \footnote{An instructive  paper by
E.~P.~Wigner, in which faster-than-light particles also have been
considered on the base of infinite-dimensional unitary irreducible
representations of the Poincar\'{e} group, was published in the 1963
\cite{wigner2}.}.
%, though has went practically unnoticed.}. 
Initially, the tachyon hypothesis encountered strong objections 
related to the principle of causality. However, it soon became clear 
that these objections are not valid if the existence of a preferred 
reference frame is assumed \cite{pdg} (called the co-moving
frame in relativistic cosmology), in which the distribution of the matter
in the universe and the relic microwave background is isotropic, and which
determines an ``absolute rest frame". In this frame 
the propagation of tachyons is ordered by retarded causality 
\cite{sigal,pvmich,pvcaus}, i.e. that effects always follow their causes,
and no causal paradoxes related to tachyons  appear either in this frame 
or, as a consequence of the theory of relativity, in any other 
reference frame \cite{rembl,radzik,ttheor,causal}.

If tachyons carry electric charge they are supposed to radiate Cherenkov photons
\cite{bds}, the Cherenkov cone angle $\theta_c$ being related to the
tachyon velocity $\beta$ in the same way as for ordinary particles:
\begin{equation}
\cos \theta_c = \frac{1} {n \beta}~,
\end{equation}
where $n$ is the refraction index of the medium traversed by the tachyon.

For the full description of how the anomalous rings were searched for, 
the reader is referred to Paper~1, which includes a description of the
DELPHI experiment and its Barrel RICH detector, the event topologies
studied and the cuts used to select the events, systematic effects
producing hit patterns imitating anomalous rings and a detailed investigation
and discussion of the different backgrounds that could fortuitously lead
to the reconstruction of anomalous rings as well as the algorithm used to
search for the anomalous rings. Sections 2$-$5 in the present paper 
represent an abbreviated version of the full description in Paper~1 
with the purpose to provide, in the present paper, a condensed background
to the interpretation given in Section~6 of the observed anomalous rings as
produced by tachyons. 
%This organized as follows. In Section~2 we describe the experimental 
%method and present several examples of anomalous ring candidates. 
%The topologies of events investigated are described in Section~3. 
%Section~4 itemises the selection cuts used in a preliminary selection
%of events of the above topologies that are subsequently investigated for the
%presence of anomalous rings; it terminates with a description of the criteria
%of the final selection of these rings. 
%Section~5 demonstrates the agreement between
%radii of anomalous rings found in the liquid and gaseous radiators of the RICH.
%A kinematic treatment of the events which contain anomalous rings
%%, including the outcome of multiple-constrained kinematic fits of these events,
%is described in Section~6. 
%The main results of this work are presented in Section~7
%and discussed in Section~8.
A summary and conclusions are given in Sections~7 and 8. 

There is an Appendix to the current paper
%which is complementary to the main text.
which lists the constraint equations used in the kinematic fitting of the
events containing anomalous rings, assuming their tachyon origin.   

%In order to keep the integrity of the presentation, Sects.~2, 4 and~5
%of the current paper reproduce Sects.~2, 4 and 10 of Paper I,
%while for getting the description of the ring finding algorithm and details of 
%background studies the use of that paper is proposed.
%%; we apologize for this inconvenience.

\section{Experimental technique}
%\footnote{This section is a revised version of Sect.~2 of Paper~I.} }
The DELPHI detector is described in detail in \cite{delphi1,delphi2}.
The following is a list of the subdetector units relevant for
this analysis: the vertex and inner detectors (VD and ID), the main tracker
%of the DELPHI detector called
(the Time Projection Chamber, TPC), the outer
detector (OD), the barrel electromagnetic calorimeter (the High-density
Projection Chamber, HPC), the hadronic calorimeter (HCAL), and the barrel muon
chambers (MUB). In some cases the information from 
%the forward electromagnetic calorimeter (FEMC) and 
the small angle tile calorimeter (STIC) was also used. 

The principal detector used in this analysis was the barrel
Ring Imaging Cherenkov detector (Barrel RICH).

The DELPHI Barrel RICH is described in detail in 
\cite{rich1,rich2,rich3,rich4,rich5}. It contained two radiators, the liquid
radiator ($C_6 F_{14}$, refraction index $n=1.273$) and the gaseous radiator
($C_5 F_{12},~n = 1.00194$). The Cherenkov photons from these two radiators 
were detected by TPC-like photon detectors consisting of 24 pairs 
of quartz drift tubes (drift boxes), covering a full azimuthal range 
with 24 Barrel RICH sectors, extended to $\pm 155$~cm from the mid-wall. 
The drift gas (75\% methane $CH_4$ and 25\% ethane $C_2 H_6$) was doped
with 0.1\% of the photosensitive agent TMAE, by which the ultraviolet
Cherenkov photons were converted into single free photoelectrons, the mean
photon conversion length being equal to 1.8~cm. The TMAE photo-ionization
threshold of 5.63~eV and the transmission cut-off of the liquid radiator and
drift tube quartz windows of 7.50~eV limited the photon wavelength detection
range to the band of 165 - 220~nm.

The principles of the creation of Cherenkov ring images in both radiators of 
the DELPHI Barrel RICH are illustrated in the following figure:
\includegraphics[height=8.0cm,width=18.0cm]{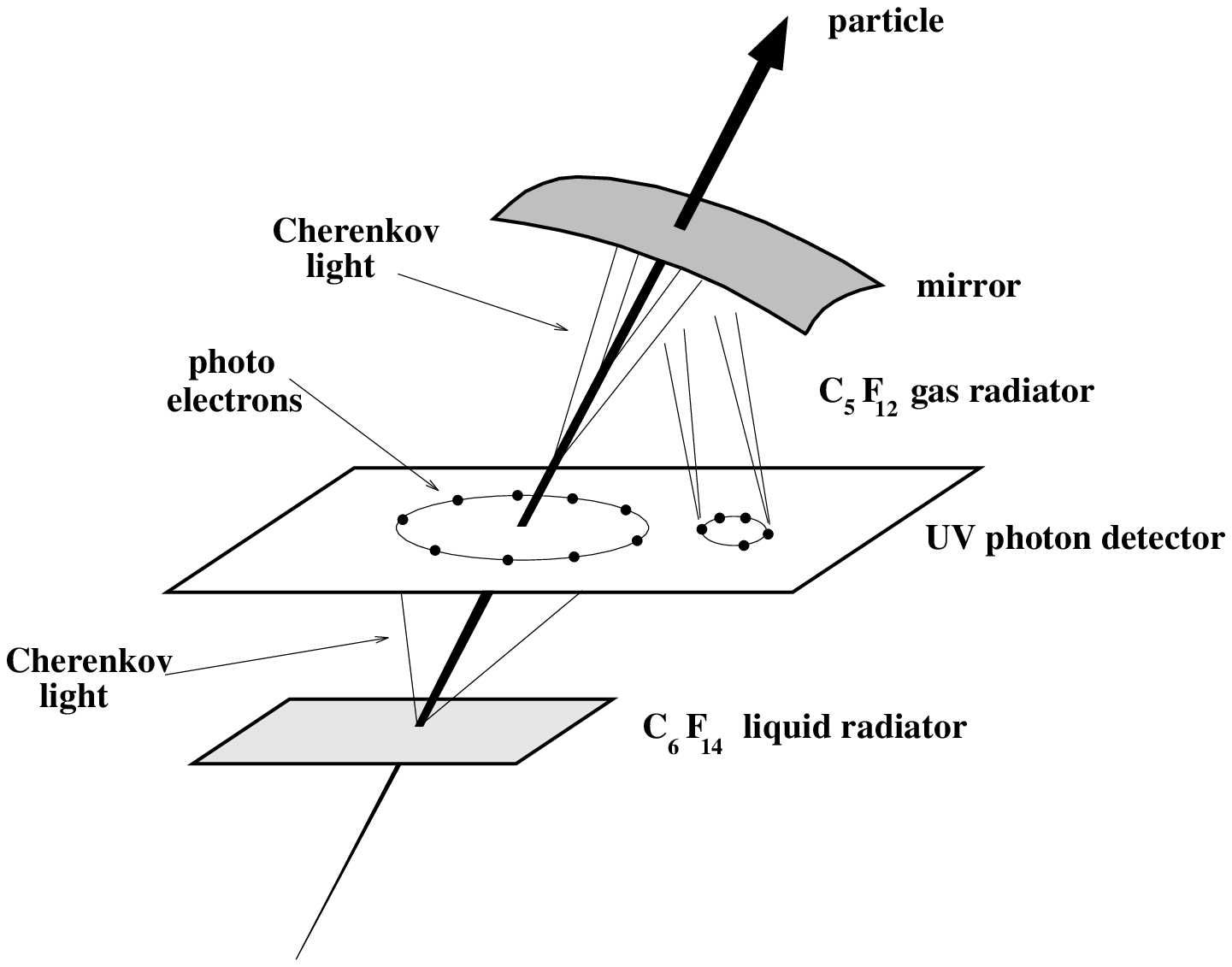}
The rings formed by the hits of photons originating from a given radiator were 
distinguished by the location of the hits in two different parts of the drift 
boxes, separated by the box median plane, the inner part for the liquid and 
outer part for the gaseous radiators (in the present paper ``inner" and 
``outer" positions in the DELPHI detector is with respect to the 
colliding beam axis).

The standard ring radii in the liquid and gas radiators, expressed in angular 
units, were 667 and 62~mrad, respectively. The total single photon standard
error of the liquid radiator $\sigma_{p.e}$ was 18-28~mrad
for standard rings. Each sector of the gaseous radiator was equipped
with six parabolic mirrors distributed along the $z$ axis~\footnote{In the 
DELPHI reference frame the $z$ axis is along the direction of the $e^-$ beam.
It defined the particle polar angles $\Theta$, while the particle azimuthal
angles $\Phi$ were defined in the $xy$ plane.},
which focused the Cherenkov photons generated in the radiator volume back onto
the photon detector. The gas radiator single photon standard error  
$\sigma_{p.e.}$ was about 4~mrad for small radius anomalous rings, increasing 
by an order of magnitude for very big rings due to geometric aberrations 
intrinsic to parabolic mirror optics. With the average Cherenkov photon numbers
per standard ring of 14 and 8~\cite{delphi2}~\footnote{These numbers reduce to 
9 and 5, respectively, within $\pm 1 \sigma_{p.e}$ cited above.} 
for the liquid and gaseous radiators, respectively, the Cherenkov cone angular 
error $\sigma$ for the rings associated with high momentum hadronic tracks
was about 9~mrad for the liquid radiator and about 2.7~mrad for the gaseous 
radiator \cite{kluit}.

After the processing of the raw data with the use of the DELPHI general pattern
recognition package $DELANA$ \cite{delphi2} to produce the DELPHI data summary
tapes (DST's), the RICH data for each individual track consisted of
%two sets of
photon trajectories starting on track segments inside a given radiator
volume, and terminating inside the RICH drift boxes on hits detected in the
latter. This information was treated rather differently
in the standard DELPHI analysis and in the analysis presented here.

The DELPHI standard way was to use the reconstructed Cherenkov angles 
to identify the particle producing the track by applying the maximum 
likelihood technique \cite{delphi2}. Five mass hypotheses 
($m_e,~m_\mu,~m_\pi,~m_K,~m_p$) 
were tried and the results were used to identify the particle. 
Additionally, an independent estimation of the ring radii relevant to each 
charged particle traversing the RICH was made and the results were stored on
the DST's~\cite{kluit}, including the averaged Cherenkov angle, the number of 
photoelectrons in the ring, an estimation of the background, etc., 
together with information on individual hits, and these data
were used in this analysis as described below.  Unfortunately, 
for the present investigation, the hits which could correspond to
Cherenkov angles exceeding certain limits were not retained in this procedure,
these limits being 750~mrad for the liquid radiator and 102~mrad for the 
gaseous one.
The information about these large angle hits is not available 
on any of the types of DELPHI DST's. This obviously restricted the power of the
preliminary selection of events containing anomalous rings at the DST level.
Such a selection, described in Sect.~4, was based on other event signatures,
though the information about Cherenkov angles within the limits mentioned above
was used in part for the selection, as will be further explained.

The events tagged by this preliminary selection were reprocessed, and the RICH 
data in them were treated following a procedure specially developed for the 
present analysis. It starts with an analysis of the information stored in raw 
data tapes extracted by the DELPHI event server. Then several steps of the 
standard procedure, described above, are performed, to reconstruct the 
Cherenkov photon directions with respect to tracks,
however without any restrictions on the radiation angles. After this
the photon directions were projected, as described below, onto the plane
perpendicular to the track trajectories in the corresponding radiator which 
will here be called the Cherenkov plane. The coordinate frame in this plane has
its origin at the position of the center of the expected ring as reconstructed 
from the track parameters. The plane coordinates $x, y$ of the photon 
projections, expressed in angular units, are calculated from
the polar and azimuthal angles of the Cherenkov photons, $\theta_c$
and $\phi_c$: $x = \theta_c \sin \phi_c$, $y = \theta_c \cos \phi_c$. 
The resulting hit pattern, which as a rule is contaminated by random 
background hits, constitutes the input to the search for Cherenkov rings.
The ring search algorithm combined with the method of evaluation of the
probability that the rings are due to background fluctuations, that 
fortuitously lead to the reconstruction of ring images, is described in 
\cite{prashort}. Only the rings having the background probability below 
10\% were kept for further consideration.

In order to illustrate the above procedure several anomalous rings 
%and their radial distributions 
found with it are shown in Figs.~1-6.

\section{Event topologies studied}
\setcounter{equation}{0}
\renewcommand{\theequation}{3.\arabic{equation}}
In selecting the events for our analysis we followed the predictions
for tachyon behaviour in an experimental set up \cite{thexp} based on expected
tachyon characteristics \cite{ttheor}. In particular, it is important to
account for selection rules based on the angular momentum conservation as they
are predicted to affect the production and propagation (through the medium) of
tachyons with non-zero spin (helicity), while scalar tachyons are expected to
be excluded on the theoretical ground \cite{ttheor}.
%,unita2}.
\vskip0.2mm
{\em Topology 1. Coupled tachyon-antitachyon pair production associated with
 a high energy photon}. The simplest topology to produce non-zero helicity
tachyons in $e^+ e^-$ interactions, which is free of selection rules due to
angular momentum conservation 
%(see \cite{ttheor}), 
seems to be the reaction
\begin{equation} 
e^+ e^- \rightarrow \gamma ~t^+ t^- ,
\end{equation} 
where $\gamma$ denotes a high energy photon and $t^+, t^-$ stand for a pair of
charged tachyons, produced with a vanishingly small opening angle,
the photon and the pair going in opposite directions.
Whatever would be (non-zero) helicities of the tachyon and the antitachyon, 
their angular momenta can compensate each other, and no appreciable suppression
of the tachyon production by the selection rules mentioned above
can be expected in this case.
Schematically, this topology can be presented by a diagram shown in 
Fig.~\ref{fig:6}a (the dashed lines in it represent the particles which go
undetected; in the case under consideration these are beam particles).
In the case when two charged tachyons are non-resolved by the TPC
the common tachyon track can be distinguished by anomalously high ionization
loss ranging from single to double electron ionization response,
with the most probable value being that of double ionization. This
expectation follows from kinematic considerations: in order to produce
an identifiable intermediate ionization response at least one of two tachyons
has to possess a velocity exceeding, say, $1.1 c$. This is not very probable
if the tachyon mass parameters are of the order of several GeV/$c^2$ or less.
The typical particle energies at LEP in the events of low multiplicity are
higher than 10~GeV, and being shared by tachyons they would ensure
tachyon velocities very close to $c$. 
%However, in the particular case of
%$v > 1.1 c$ one can expect, in addition to two anomalous rings in
%the gas radiator of the RICH associated with the common tachyon-antitachyon
%track, an associated anomalous ring in the liquid radiator. Such events,
%possessing two additional tachyonic signatures (intermediate ionization
%response and/or an anomalous ring in the liquid radiator), would be
%strong tachyon candidates.

Two respective examples of topology 1 candidate events are given in 
Figs.~\ref{fig:61}-\ref{fig:161a}.\\
\vskip0.2mm
{\em Topology 2. Back-to-back (or quasi-back-to-back) tachyon-antitachyon
production.} Under certain conditions a tachyon-antitachyon pair can be
produced alone in an $e^+ e^-$~collision:
\begin{equation}
e^+ e^- \rightarrow t^+ t^- . 
\end{equation}
with the two tachyons going in opposite directions. Such a reaction is allowed
only if the tachyons have the minimal possible helicity values, 1/2
(excluding the possibility of scalar charged tachyons on theoretical grounds, 
as mentioned above).
The reaction diagram of this process is presented in Fig.~\ref{fig:6}b.
Such a process is, from the point of view of
the angular momentum conservation, analogous to an anti-neutrino scattering
on an electron. The distribution of the production angle of tachyon pair
$\Theta$ is expected in this case to be proportional to $\cos^2 \Theta$,
i.e. rather suppressed in the barrel region, vanishing at $\Theta = 90^\circ$.

The situation with the angular momentum conservation can be softened
in the case of tachyon pair production in $\gamma \gamma$ interactions,
i.e. in reaction
\vskip 0.5mm
\begin{equation}
e^+ e^- \rightarrow e^+ e^- ~t^+ t^- . 
\end{equation}
\vskip 0.5mm
\noindent
in which tachyon configuration can be defined as quasi-back-to-back one
(the final faster-than-light particles are acollinear but coplanar).
The diagram of this reaction is shown in Fig.~\ref{fig:6}c.
As can be seen comparing this diagram with the diagram of reaction (3.2),
these two reactions, in spite of being quite different in terms of the
tachyon production mechanism, generate the same
two-track, (quasi-) back-to-back topology. This topology is
similar to the Bhabha-like event topology in the case of reaction (3.2),
or to the $e^+ e^-$ pair production via $\gamma \gamma$
interaction in the case of reaction (3.3). Thus the Bhabha-like and
$\gamma \gamma$ events are expected to be the main source of background events
in this channel. The decisive signature of the events of reactions (3.2) and
(3.3) has to be the observation of anomalous rings in the RICH detector
associated with both candidates for tachyon tracks, and a successful passage
of an appropriate kinematic treatment by the events.
Two respective examples of topology 2 event candidates are shown in 
Figs.~\ref{fig:62a}-\ref{fig:162}.\\
\vskip0.2mm
{\em Topology 3. Electroproduction of a tachyon-antitachyon pair in front of
the TPC.} The suppression of the tachyon production by selection rules due to
the angular momentum conservation can be essentially weakened
if a tachyon-antitachyon pair is created in matter, in which case some
additional angular momentum degrees of freedom can appear.
Such a process could be induced, for example, by secondary electrons
(or positrons), generated in the beam-beam or $\gamma \gamma$ interactions,
when these particles pass through the detector material, with the following
reaction taking place:
\begin{equation}
e X \rightarrow  X^\prime e~ t^+ t^-, 
\end{equation}
where $X$ and $X^\prime$ denote a charged object (most probably, a nucleus
of the detector medium with which the incident electron interacts
\footnote{Due to a big tachyon longitudinal extension, which accordingly to
\cite{ttheor} should essentially exceed a nuclear scale, the coherent
electroproduction of tachyons on nuclei is expected to dominate this
process.}), before and after the interaction. 
Excepting this charged object, there are no 
undetectable charged particles in this reaction (see the reaction diagram in 
Fig.~\ref{fig:6}d). Thus a typical topology of events of this kind would be 
a presence of a ``jet" consisting of three tracks, of which at least two 
have non-zero impact parameters with respect to the primary vertex.
The particle producing the three-particle jet is expected to be opposed by 
another track in the opposite hemisphere, as shown in an example of 
this topology event candidate in Figs.~\ref{fig:63}, \ref{fig:163}.

\section{Selection of events having candidate anomalous rings}
%\footnote{This section is a revised version of Sect.~4 of Paper~I.} }
The following selection cuts were applied for the preliminary selection of
events from the DELPHI DST's:

{\em The general selection of events} was done vetoing the hadronic events
defined according to the DELPHI ``Team~4" criteria \cite{delphi2} \footnote
{The main criterion for selecting hadronic events was the requirement of
a multiplicity above 4 of charged particles with $p > 400$~MeV/$c$,
$20^\circ < \theta < 160^\circ$ and a track length of at least 30~cm
in the TPC, with a total energy in these charged particles above
$0.12 \times E_{cm}$.}. This retained about $10^7$~ events
(called ``quasi-leptonic events"), found in the 1992 - 2000 data~\footnote
{The transition from LEP1 to LEP2 took place in the year 1995.
LEP1 beam energies were centered around the mass of $Z^0$-boson
(91.2 GeV/$c^2$), the LEP2 period was aimed at physics beyond $Z^0$, with
beam energies spanning between 161 and 208~GeV (the averaged energy being
196~GeV).}. A further reduction of the amount of hadronic events was
performed by using the DELPHI electromagnetic calorimeter as described below.
Muons were identified (and rejected) with the HPC, HCAL and MUB responses.
%As a next step of the selection, all charged tracks in 
%these events were required to be neither hadrons (as identified with the HPC 
%and HCAL responses) nor muons (as identified with the HPC, HCAL and MUB 
%responses). 
Another general selection criterion was the requirement that for all tracks
of the events the errors of the momenta derived from the curvature of
the tracks be  below certain limits
%($\delta p/p < 0.05$ for low momentum tracks and 
($\delta p/p < 0.3$).
%for tracks having momenta greater than 10~GeV/$c$). 
The tracks were furthermore required to be within
the geometrical acceptance of the gas radiator of the DELPHI Barrel RICH
($46^\circ < \Theta < 134^\circ$), with the barrel RICH being in
an operational state.

The further selection of events
used the cuts designed to enhance the contents of events of topologies 1, 2
and~3 which are described below (the term $jet$ here corresponds to a single
neutral or charged particle or two or three tightly bunched charged particles),
and the results of the application of these selections to the DELPHI DST data
is shown in Tables~1 and~2 of Paper I.

\subsection{Topology 1}
\begin{itemize}
 \item [1.] Two jets in the opposite hemispheres are required in the event:
one neutral jet and one jet consisting of 1 or 2 charged particles; each
of the jets should have at least 50\% of the beam energy.
%; in the case of the neutral jet this energy should come from the HPC. 
 \item [2.] Track(s) of the charged jet should have associated shower(s) in
the HPC with the total jet shower energy exceeding half of the jet momentum,
the number of active HPC layers in the shower(s) being greater than or equal
to 5 (of 9).
 \item [3.] Track(s) of the charged jet should have the first measured point(s)
in the first layer of the DELPHI vertex detector (VD).
 \item [4.] In the case of a single-track charged jet (topology 1a) the track
ionization has to be within the limits of $2.0 < dE/dx < 3.8$ mips~\footnote
{A mip is the $dE/dx$ of a minimum ionization particle.} 
(corresponding to the 2 charged tracks being non-separable in the TPC),
with the number of TPC wires available for ionization measurement exceeding 40
out of a total of 192 wires (this cut aimed at the suppression of Compton event
background).
 \item [5.] In the case of a jet with two separated tracks (topology 1b)
the jet mass, calculated under hypothesis of the electron mass for the tracks,
has to be below 0.5~GeV/$c^2$, and the smaller of the two track momenta
has to exceed 4~GeV/$c$~\footnote{This cut was applied in order to keep the
efficiency of the track association with its HPC shower close to 100\%.}.
To suppress the Compton event background the jet acollinearity was required
to be below $2^\circ$.
\end{itemize}

It is worth to note that for a proper finding of Cherenkov rings (standard
or anomalous ones) only the track direction in the RICH, and not its
momentum value, is required.
This direction is considered to be well defined even for the events in which
two tracks were not separated (not resolved) in the TPC, identified by
double ionization only (the tracks from point~4 above). The reason for
that these tracks were not resolved in the TPC was,
in an addition to their negligible opening angle,
their high momenta leading to a small curvature of their trajectories
in the DELPHI detector magnetic field. This resulted in the appearance of
these tracks as practically straight lines in the detector; these tracks
produced two Cherenkov rings in each of the RICH (liquid and gaseous)
radiators (in the case of electron tracks all the rings are standard).

\subsection{Topology 2}
\begin{itemize}
 \item [1.] Two tracks of opposite charge are required in a 2-jet event, one
track per jet, both tracks having momenta greater than 4~GeV/$c$.
Each jet should have at least 60\% of the beam energy (topology 2a, Bhabha-like
events) or, alternatively, each jet should have less than 60\% of the beam
energy (topology 2b, $\gamma-\gamma$ events).
 \item [2.] Both tracks of the event should have associated showers in
the HPC with each shower energy exceeding half of the track momentum,
the number of active HPC layers in the showers being greater than or
equal to 5 (of 9).
 \item [3.] The tracks have to go in opposite directions in the $xy$ plane
having the track acollinearity lower than $2^\circ$ (in the case of
topology~2a) or, in the case of topology 2b, having the track acollinearity
higher than $2^\circ$ and the track acoplanarity below~$4^\circ$.
 \item [4.] Each of the tracks should have a gas Cherenkov angle,
%as calculated using the hits in the gas radiator of the Barrel RICH and 
stored on the DST, within the region 72 to 102~mrad~\footnote{The upper limit
is defined by a restriction on the gas Cherenkov radii imposed when producing
the DST's.}, with at least 5 photoelectrons associated with each ring;
 \item [5.] or, alternatively, one of the two tracks should have a
Cherenkov angle within the region 80 to 102~mrad,
with at least 5 associated photoelectrons.
\end{itemize}

\subsection{Topology 3}
\begin{itemize}
 \item [1.] Two jets are required in the event: one jet consisting of a single
track, and the other consisting of 3 charged particles, with the overall sum
of the track charges equal to zero. All the tracks in the 3-particle jet should
have momenta exceeding 3~GeV/$c$.
\item [2.] Tracks in the 3-particle jet should have associated showers in the
HPC with each shower energy exceeding half of the track momentum, the number of
active HPC layers in the showers being greater than or equal to 5 (of 9).
\item [3.] For a primary suppression of $\tau-\tau$ events (abundant in LEP1
data set) the effective mass of charged particles in the 3-particle jet
(calculated prescribing pion masses to the tracks) is demanded to be greater
than 0.5~$m_{\tau}$ (0.89~GeV/$c^2$), and the jet momentum is required to be
greater than 0.8 of the beam momentum.
\item [4.] Two tracks in the 3-particle jet are required to
have the first measured point at the detector radius $R > 35$~cm (i.e. outside
the ID, the DELPHI Inner Detector). And at least one track in this jet should
have an impact parameter with respect to the primary vertex in the $xy$ plane
exceeding 6~mm, or two tracks should have impact parameters each exceeding 4~mm,
while the impact parameters of all tracks with respect to the primary vertex
should be below 10~cm.
 \item [5.] At least two tracks in the three-track jet should have a
non-standard RICH response from the gas radiator (stored on the DST), i.e.
either ring radii outside of the standard ring limits of $52 < r < 72$~mrad,
or no response at all \footnote{This non-standard response was expected in the
cases of anomalous rings having radii greater than 102~mrad.},
while each track is required to have some response from the liquid
radiator; these tracks should be within the angular acceptance of the gas
radiator i.e. within the range of $46^\circ < \Theta < 134^\circ$.
% \item [5.] Only one gas standard ring of radius from 57 to 67 mrad found by
%the computer-coded algorithm having the background probability less than 
%10\% and containing more than 3 hits is allowed in the 3-particle jet.
\end{itemize}

\subsection{Final selection}
The cuts described in Sects. 4.1 to 4.3 resulted in the selection
of 395 events as primary signal candidate events of topologies 1, 2 and 3
(which correspond to the sum of events in the Selection~4 row of Tables~1
and 2 in Paper I.

A further analysis of all the selected events was done retrieving these events
from the DELPHI raw data using the DELPHI event server, reconstructing
the Barrel RICH hit patterns for all tracks in these events and searching for
anomalous rings associated to each track, as described in Appendix~1 
of Paper I.

Of the 395 selected events 53~events were found to
have at least one anomalous ring with a probability to be composed
of background hits below 10\%; the distribution of these events
over different topologies is shown in Selection 5 rows of Tables 1 and 2 
of Paper~I. The minimal number of photons per ring was required to be 4.
In addition, the events of topologies~1 and~2 were required not to have
standard rings in the gas radiator.

Of these 53 events 29 events were found to possess two gaseous anomalous rings
(at least one anomalous ring per track in topology 2). All these events
were subjected to a further treatment, aimed at the finding of rings,
standard and/or anomalous, in the liquid radiator.
The search for liquid radiator rings in the events with the found gas rings,
combined with the method of evaluation of the probability
that the ring is composed of background hits, was performed
as described in Appendix 2 of Paper I. The range of the liquid ring radii
searched for with this method spanned between 600 and 1100~mrad.
In the case of such a ring being found with the ring radius exceeding 700~mrad
(excluding quartz ring range, 857 - 909 mrad)
it was considered as an anomalous liquid radiator ring; only those anomalous
liquid radiator rings having the background probability below 10\%
were accepted for further analysis.
Then the background probabilities of all anomalous rings in a given event
were multiplied, and the product of the probabilities was required to be less
than $10^{-3}$. Of the selected 29 events this left 9 events of topology 1,
6 events of topology 2, and 12~events of topology 3, in total 27 events
as given in the Selection~6 rows in Tables 1 and~2 of Paper I. None of the 
anomalous rings in these events were compatible with being due to the systematic 
effects described in Section~5 of Paper I. The characteristics of the rings 
contained in all the 27 selected events are given in Tables 3 to 5 of Paper~I. 
%In total, the number of selected events with at least two anomalous rings
%reduced to 27/17/10/6 when the combined probability that the reconstructed
%rings are due to background, mimicking ring images, is below
%$10^{-3}/10^{-4}/10^{-5}/10^{-6}$, respectively. 

The studies of backgrounds and possible systematic effects, which can
lead to the appearance of spurious anomalous rings, are described in detail 
in Sects.~6 to 9 of Paper I. It follows from these studies that the probability 
that the events with the observed anomalous rings can be explained by known 
physical processes is very low (below $10^{-9}$). Moreover, the comparison of 
the rates of events with single and double anomalous rings, carried out 
in Sect.~8 of Paper I, indicated a clear tendency for the observed anomalous 
rings to be produced in pairs.

\section{Correlation between gaseous and liquid radiator rings} 
%\footnote{This section is a revised version of Sect.~10 of Paper I.}}
\setcounter{equation}{0}
\renewcommand{\theequation}{5.\arabic{equation}}
An important feature of the DELPHI RICH was the presence
of two radiators: the liquid radiator and outside it, the gaseous one.
With the use of the relationship originating from Cherenkov formula
\begin{equation}
  n_{liq} \cos{\theta_{liq}} = n_{gas} \cos{\theta_{gas}},
\end{equation}
the ring radius expected in the liquid radiator can be derived from the
gaseous ring radius and {\em vice versa}. 

Fig.~\ref{fig:15} shows the anomalous ring radii observed in the gas radiator
plotted against the ring radii observed for the same track
in the liquid radiator for 27 events listed in Tables~1, 2 and~3 of Paper~I,
showing a clear correlation between the radii.
The curved line in the figure is not a fit to the data, but has been derived
using the relation (5.1) with $n_{liq} = 1.273$ and $n_{gas} = 1.00194$.

The same correlation, presented in terms of particle velocities as calculated
from the ring radii in both radiators according to formula (1.1),
$\beta_{liq}$ and $\beta_{gas}$, is shown in Fig.~\ref{fig:16}.
The correlation coefficient calculated for 53 points of
Fig.~\ref{fig:16} equals 0.992. The probability of obtaining such a coefficient
for uncorrelated variables was estimated on the basis of an
analysis of $10^9$ MC ``toy experiments" in each of which the values of
the 53~velocity pairs, $\beta_{liq}$ and $\beta_{gas}$,
were generated as distributed randomly according to the
corresponding projections of the plot in Fig.~\ref{fig:16}. The probability
resulting from this analysis is $10^{-9}$.

The sum of $\chi^2$ for deviations of 53 points in Fig.\ref{fig:16} from the 
main diagonal is 40.1. The corresponding probability for that these points are 
in agreement with the predictions of formula (5.1) is near 95\%.  
 
%The accumulation of points in the left bottom corner of Fig.~\ref{fig:15}
%is explained by the fact that of the 53 gas rings, represented in 
%Fig.~\ref{fig:15}, as many as 35 rings have a radius less than 120~mrad.
%In particular, all the events of topology~2 are concentrated
%in this corner owing to the selection cuts used, see Sect.~4.2 and Table~4:
%the anomalous gas rings with the radii below 120~mrad
%referred to in this Table
%correspond to the liquid radiator ring radii below 673~mrad, which are
%very close to the standard liquid ring radii $667 \pm 9$~mrad;
%a similar conclusion is applicable also for anomalous gas rings with radii
%below 200~mrad in the events of other topologies.

\section{Kinematic treatment of events containing anomalous rings}
\setcounter{equation}{0}
\renewcommand{\theequation}{6.\arabic{equation}}
This analysis consisted of the following two-step procedure.

First, the momentum $p$ of a tachyon is derived from the information 
provided by the DELPHI tracking devices. Since a tachyon is a
space-like particle, the square of its 4-momentum $P$, 
$P^2 = E^2 - p^2 =-\mu^2$, is negative, where $E$ is the tachyon energy
and $\mu$ is the tachyon mass parameter. The velocity of the tachyon is defined
as for ordinary particles
to be $\beta = p/E, \beta > 1.$ Using the measured Cherenkov angle $\theta_c$ to
determine the velocity from relation $\beta = 1 / n~\cos\theta_c$ one obtains
the following expression for the tachyon mass parameter:
\begin{equation}
\mu = \frac{p}{c} \sqrt{1 - (n~\cos\theta_c)^2}  
\end{equation}
where $n$ is the refraction index of the corresponding radiator. The
distributions of mass parameters calculated with formula (6.1)
for tachyon track candidates
are shown in Fig.~\ref{fig:17}a, b, separately for regions of low and high
mass parameter values. One can see that there are two clusters of entries
grouping around $0.28$~GeV/$c^2$ (Fig.~\ref{fig:17}a) and $5$~GeV/$c^2$
(Fig.~\ref{fig:17}b).

Then for those events which have both momenta of tachyon track candidates
reconstructed by DELANA (always of opposite electric charge) we plot,
in Figs.\ref{fig:17}c,~\ref{fig:17}d, a two-dimensional distributions of 
mass parameters of the tachyon track candidates.
One can see that the two clusters of entries mentioned above can be seen 
in this figure are present also, having the cluster centers close to 
the diagonal of the plot. Moreover, a third cluster consisting of 5 events
around 1 GeV/$c^2$ can be seen in this plot. The correlation coefficient
calculated for 19 points on the plot in Fig.~\ref{fig:17}d
equals to 0.970. The probability of obtaining such a high coefficient for
uncorrelated mass parameters was estimated to be below $        10^{-6}$.
This estimation is based on the analysis of $10^7$ MC ``toy experiments",
in each of which the values of 19~mass-parameter pairs were generated
as distributed randomly within the ranges of the axes of
Fig.~\ref{fig:17}d.

This result is interpreted as an indication for mass parameters 
of the tachyon and the antitachyon candidates produced in a given
event to be of equal value. This indication, together with the formal 
arguments in \cite{wigner2} for the tachyon and antitachyon mass parameters
to be equal, was used as an argument to keep the mass parameter values
equal in the further
kinematic treatment of each of selected tachyon candidate event.

This further treatment was based on an over-constrained kinematic fit
of the event variables, i.e. with the number of
constraint equations exceeding the number of unknowns. 
These equations are listed in the Appendix of this paper.  
Four equations based on energy-momentum conservation (equations A.1-4
and A2.1-4 in the Appendix) plus two or more
equations based on  Eq.~6.1 (equations A1.5-6 and A2.5-6
in the Appendix), one for each anomalous ring 
found in a given event, were used in the fit. 
In the events of topology 1a, in which tachyon track candidates 
were not resolved in the TPC, an additional constraint equation was implied 
using the $dE/dx$ information from the common tachyon-antitachyon track
candidate (equation A1.7 in the Appendix $-$ for this is used that the $dE/dx$
for tachyons is assumed to be equal to that the Fermi plateau for
$\beta\rightarrow 1$ and decrease with increasing $\beta$ above 1 
like $1/\beta^2$, see ref. 21, p.~6).

The unknowns were the tachyon mass parameter $\mu$,
assumed to be equal for the tachyon and antitachyon, and for topology 1a
the energy of the tachyon whose momentum was not measured. In the
events of this topology and of topology 1b an additional unknown, the energy 
of an initial state radiation (ISR) photon, was treated in the case of 
the acollinearity of the tachyon-antitachyon candidate pair 
with the high energy photon detected in 
a given event exceeded $1^\circ$ (which would correspond to the ISR photon 
energy exceeding 1.5~GeV). In the events of topology 2, reaction (3.3), 
instead of this unknown two other additional unknowns were treated, the 
algebraic sum of momenta of the final state $e^+ e^-$ and their invariant mass.

The events of topology~3 41633:1568, 104949:27827 and 105892:13127, having 
visible energy compatible with the beam $e^+ e^-$ pair $\sqrt s$ and the jet 
acollinearity below $1^\circ$, were treated as events of topology 1 (without 
the ISR photon) \footnote{There was an exceptional event (114204:2668) of 
this topology in which both tachyon candidate tracks were badly measured, which
together with other kinematic uncertainties of this event prevented a kinematic 
fit. The event 79940:2890 was fitted with 1 kinematic equation
(instead of 4), using the energy conservation for the 3-particle jet
($e^- t^+ t^-$), with the jet shower energy measured by the HPC.}.
In the remaining events of this topology the over-constrained
kinematic fit was built over constraint equations using the Cherenkov angles
of the anomalous rings found in these events, the number of unknowns being one,
namely, the tachyon mass parameter (common for both tachyon track candidates).   
The fit results are given in Tables 1 -3, together with some input 
kinematic parameters, including measured radii of the anomalous Cherenkov
rings and their errors. Also the number of degrees of freedom of the kinematic 
fit in a given event is shown. We give no post-fit errors of the measured
parameters in the tables since they are in most cases practically equal to
the input errors, only slightly decreased by the fit.

%All events in these Tables were accepted as tachyon event candidates
%having the fit $\chi^2$ per number of degrees of freedom below 4.
The distributions of fitted mass parameters are shown in Fig.~\ref{fig:18}.
The peaks in these distributions were fitted with Gaussians. In the low
mass region the Gaussian mean value equals to $(0.290 \pm 0.005)$~GeV/$c^2$
with the standard deviation (width) of $(0.012 \pm 0.004)$~GeV/$c^2$.
In the high mass region the Gaussian mean value equals to
$(4.55\pm 0.12)$~GeV/$c^2$ with the standard deviation of
$(0.40 \pm 0.12)$~GeV/$c^2$.
In the both cases the peak widths are compatible with the accuracy of the
determination of the tachyon mass parameter in the appropriate mass region.
Note also the cluster of 5 events spread about the tachyon mass 
parameter of 0.9~GeV/$c^2$.
 
\section{Summary}
\begin{itemize}
\item [1).] The observation of particle trajectories producing Cherenkov
rings with radii significantly exceeding the standard ones (produced by
ultrarelativistic particles) is interpreted as evidence for the existence
of tachyons, i.e. particles travelling faster than light.
\item [2).] The studies of backgrounds and possible systematic effects,
which can lead to the appearance of spurious anomalous rings, were carried out 
as described in Paper I and have shown that the probability that the events 
with the observed anomalous rings can be explained by known physical processes 
is very low ($10^{-3}$ or less). Moreover, a comparison of rates 
with a single and double gaseous anomalous rings indicates a clear tendency 
for the observed anomalous rings to be produced in pairs.   
\item [3).] In all the cases when anomalous rings associated with a given
track are observed in both the liquid and the gaseous radiators, the ring
radii are highly correlated according to the respective refractive indices of 
these radiators (see formula (5.1) and Figs.~\ref{fig:15} and~\ref{fig:16}). 
The correlation coefficient for the points in Fig.~\ref{fig:16} is high 
(0.992) and the probability to obtain such a correlation coefficient for 
uncorrelated data is low (estimated to be below $10^{-9}$).
\item [4).] For tachyon track candidates, associated with anomalous rings
and  having well measured momenta, the tachyon mass parameter can be calculated
using formula~(6.1). The calculated masses group around 0.28 and
5~GeV/$c^2$ (see Fig.~\ref{fig:17}a, b). In the events when both tachyon track
candidates have measured momenta (always of opposite sign) 
the calculated masses group around the main diagonal of the 2-dimensional 
mass distributions (see plots~\ref{fig:17}c,~\ref{fig:17}d), 
which indicates the tachyon-antitachyon mass parameter equality 
predicted by the theory \cite{wigner2} (see also \cite{ttheor}).
Correlation coefficient for the points on the plot~\ref{fig:17}d
equals to 0.970. The probability of obtaining such a correlation coefficient
for uncorrelated mass parameters is below $        10^{-6}$.
\item [5).] 26 of the 27 selected events successfully passed the 
over-constrained kinematic fits based on the tachyon production hypothesis 
for these events. The distributions of the tachyon mass parameters of the 
fitted events (Fig.~\ref{fig:18}) show two peaks against the vanishing
background at tachyon mass parameters of 0.29 and 4.5~GeV/$c^2$, 
with the peak widths being compatible with the expected tachyon mass parameter
determination accuracy in the appropriate mass regions. 
\end{itemize}

\section{Conclusion}
An analysis of events with anomalous Cherenkov rings in the DELPHI Barrel RICH
has been made under the hypothesis that these rings result from the radiation
of faster-than-light particles, i.e. tachyons.

A detailed study of backgrounds carried out in the Paper I
%and possible systematic effects, capable of producing apparently anomalous rings, 
indicates that the probability that the
reconstructed rings are the result of fortuitous combinations of background
hits is low. An argument against the background hypothesis is provided
by the observation of a high correlation between anomalous ring radii
found in the liquid and gas radiators seen in Figs.~\ref{fig:15},
\ref{fig:16}.

In the framework of the tachyon interpretation of the particles associated
to the observed anomalous rings and assuming the equality of the tachyon and
antitachyon mass parameters indicated by the mass correlation plot shown in
Fig.~\ref{fig:17}d) an over-constrained kinematic fit of 26 of 27~selected
events could be tried. The distributions
of the tachyon mass parameters derived from the fit
is narrowly peaked at two mass values, at $0.290 \pm 0.005$~GeV/$c^2$
and $4.55 \pm 0.12$~GeV/$c^2$ with little background, displayed in
Fig.~\ref{fig:18}.

Our conclusion is that further searches for the faster-than-light particles
should be made in dedicated experiments
in order to corroborate or refute the findings of this analysis.

One can use for the tachyon search the existing data of
past and current experiments, e.g. the data of the ALICE experiment
at LHC which contains a RICH (called HMPID) and a TOF system, the latter
with a declared resolution of 80-120~ps \cite{alice}.
%Events containing electron-like particles can be selected for the analysis. 
The optimal processes for the investigation
could be $\gamma \gamma$ interactions (corresponding to
our topology 2, reaction 3.3), with an expected $Z^2$ enhancement factor
for the tachyon production in Pb-Pb collisions.
The data of the LHCb experiment \cite{lhcbdet} which contains RICH detectors
with a high Cherenkov angle resolution \cite{lhcb} also can be used selecting
low multiplicity events in the data of pp, pPb and Pb-Pb collisions.
\subsection*{Acknowledgments}
%\vskip 3 mm
We are greatly indebted to the DELPHI Collaboration for the permission
to use the DELPHI experiment data in our analysis.

We thank Profs. Yu. G. Abov, K.~G.~Boreskov, F.~S.~Dzheparov, O.~V.~Kancheli, 
A.~M.~Kunin and Drs. F.~Dydak, O.~N.~Ermakov, A.~I.~Golutvin, 
A.~A.~Grigoryan, S.~Sila-Novitsky for fruitful discussions and 
Drs. U.~Schwickerath and R.~M.~Shahoyan for technical assistance. 

\newpage
\section*{Appendix. Constraint equations of kinematic fits}
\subsection*{Topology 1}
~~~~~~Definitions:\\
$p_{t1}$ - measured tachyon candidate 1 track momentum;\\
$\Theta_{t1}$ - measured common tachyon candidate pair polar angle;\\
$\Phi_{t1}$ - measured common tachyon candidate pair azimuthal angle;\\
$E_{\gamma}$ - measured HE photon energy;\\
$\Theta_{\gamma}$ - measured HE photon polar angle;\\
$\Phi_{\gamma}$ - measured HE photon azimuthal angle;\\
$\Theta_{ISR}$ - ISR photon polar angle $\Theta$: measured by the STIC or, otherwise, 
put to $0^\circ$ or $180^\circ$~\footnote{The choice between $0^\circ$ and 
$180^\circ$  was definite, requiring the positivity of the ISR photon energy, 
$\tau_{ISR} > 0$ ($\tau_{ISR}$ to be defined below).};\\
$\Phi_{ISR}$ - ISR photon azimuthal angle $\Phi$: measured by the STIC or, otherwise, 
put to $0^\circ$;\\
$R_{t1}$ - measured radius of the anomalous ring associated with the tachyon 
candidate~1;\\
$R_{t2}$ - measured radius of the anomalous ring associated with the tachyon 
candidate~2;\\
$n_1$ and $n_2$ - refraction indices of the radiators in which the corresponding
rings were found;\\
$\rho$ (in the case of topology 1a) - measured common tachyon candidate track
$dE/dx$ reduced by the Fermi plateau value (1.60 mips in the DELPHI TPC);\\
$\mu$ - unknown (common) tachyon mass parameter;\\
$\tau_{ISR}$ - unknown ISR photon energy;\\
$E_{t2}$ - unknown energy of the second tachyon in the case of topology 1a;\\
$p_{t_2}$ - momentum of the second tachyon: measured in the case of topology 1b,
otherwise equal to $\sqrt{\mu^2 + E_{t2}^2}$;\\
$\epsilon_{t_2}$ - energy of the second tachyon: unknown ($E_{t2}$) in the case of
topology~1a, otherwise equal to $\sqrt{p_{t_2}^2 - \mu^2}$;\\ 
$E_{c.m.s.} = \sqrt{s}$.
\vskip1mm
Equations:
\vskip1mm
$
f_1 = (p_{t1}+p_{t_2})\sin \Theta_{t1}\cos \Phi_{t1} + E_{\gamma}\sin \Theta_{\gamma}\cos \Phi_{\gamma} + \tau_{ISR}\sin \Theta_{ISR}\cos \Phi_{ISR}
=0$~~~~~~~~~(A1.1)
\vskip1mm
$
f_2 = (p_{t1}+p_{t_2})\sin \Theta_{t1}\sin \Phi_{t1} + E_{\gamma}\sin \Theta_{\gamma}\sin \Phi_{\gamma} + \tau_{ISR}\sin \Theta_{ISR}\sin \Phi_{ISR}
=0$~~~~~~~~~(A1.2)
\vskip1mm
$
f_3 = (p_{t1}+p_{t_2})\cos \Theta_{t1}         + E_{\gamma}\cos \Theta_{\gamma}         + \tau_{ISR}\cos \Theta_{ISR}        
=0$~~~~~~~~~~~~~~~~~~~~~~~~~~~~~~~~~~~(A1.3)
\vskip0mm
$
f_4 = \sqrt{p_{t1}^2 - \mu^2} + \epsilon_{t_2}+ E_{\gamma} + \tau_{ISR} - E_{c.m.s.}
=0$~~~~~~~~~~~~~~~~~~~~~~~~~~~~~~~~~~~~~~~~~~~~~~~~~~~~~~~~~~~~~~~~~~~~(A1.4)
\vskip0mm
$
f_5 = \sqrt{p_{t1}^2 - \mu^2}/p_{t1} - n_1 \cos R_{t1}
=0$~~~~~~~~~~~~~~~~~~~~~~~~~~~~~~~~~~~~~~~~~~~~~~~~~~~~~~~~~~~~~~~~~~~~~~~~~~~~~~~(A1.5)
\vskip1mm
$
f_6 = \epsilon_{t_2}/p_{t_2} - n_2 \cos R_{t2}
=0$~~~~~~~~~~~~~~~~~~~~~~~~~~~~~~~~~~~~~~~~~~~~~~~~~~~~~~~~~~~~~~~~~~~~~~~~~~~~~~~~~~~(A1.6)
\vskip1mm
In the case of topology 1a:
\vskip1mm
$
f_7=(1/\beta_{t1}^2+1/\beta_{t2}^2)-\rho = 1-\mu^2/p_{t1}^2+\epsilon_{t_2}^2/p_{t_2}^2 - \rho=0,
$~~~~~~~~~~~~~~~~~~~~~~~~~~~~~~~~~~~~~~~~~~~~~~~(A1.7)
\vskip1mm
where $\beta_{t1}$ and $\beta_{t2}$ are velocities of the tachyons 1 and 2 defined by gas rings,
\cite{thexp}, p.~6. 

\subsection*{Topology 2}
~~~~~~Definitions:\\
$p_{t1}$ - measured tachyon candidate 1 track momentum;\\
$\Theta_{t1}$ - measured tachyon candidate 1 track polar angle;\\
$\Phi_{t1}$ - measured tachyon candidate 1 track azimuthal angle;\\
$p_{t2}$ - measured tachyon candidate 2 track momentum;\\
$\Theta_{t2}$ - measured tachyon candidate 2 track polar angle;\\
$\Phi_{t2}$ - measured tachyon candidate 2 track azimuthal angle;\\
$\Theta_{ee}$ - polar angle of the final $e^+, e^-$ summary momentum:
measured or, otherwise, put to $0^\circ$ (not necessary in the case of 
topology 2a);\\
$\Phi_{ee}$ - azimuthal angle of the final $e^+, e^-$ summary momentum:
measured or, otherwise, put to $0^\circ$ (not necessary in the case of 
topology 2a);\\
$R_{t1}$ - measured radius of the anomalous ring associated with the tachyon 
candidate 1;\\
$R_{t2}$ - measured radius of the anomalous ring associated with the tachyon 
candidate 2;\\
$n_1$ and $n_2$ - refraction indices of the radiators in which the 
corresponding rings were found;\\
$\mu$ - unknown (common) tachyon mass parameter;\\
$p_{ee}$ - the algebraic sum of the final $e^+, e^-$ momenta (zero in the case of
topology 2a);\\
$m_{ee}$ - the effective mass of the final $e^+ e^-$ pair (zero in the case of
topology 2a);\\
$E_{c.m.s} = \sqrt{s}$.

\vskip1mm
Equations:
\vskip1mm
$
f_1 = p_{t1} \sin \Theta_{t1}\cos \Phi_{t1} + p_{t2}\sin \Theta_{t2}\cos \Phi_{t2} + p_{ee}\sin \Theta_{ee}\cos \Phi_{ee}
=0$~~~~~~~~~~~~~~~~~~~(A2.1)
\vskip1mm
$
f_2 = p_{t1} \sin \Theta_{t1}\sin \Phi_{t1} + p_{t2}\sin \Theta_{t2}\sin \Phi_{t2} + p_{ee}\sin \Theta_{ee}\sin \Phi_{ee}
=0$~~~~~~~~~~~~~~~~~~~~(A2.2)
\vskip1mm
$
f_3 = p_{t1} \cos \Theta_{t1}         + p_{t2}\cos \Theta_{t2}         + p_{ee}\cos \Theta_{ee}        
=0$~~~~~~~~~~~~~~~~~~~~~~~~~~~~~~~~~~~~~~~~~~~~~~~~~~~~~~(A2.3)
\vskip0mm
$
f_4 = \sqrt{p_{t1}^2-\mu^2} + \sqrt{p_{t2}^2-\mu^2} + \sqrt{p_{ee}^2+m_{ee}^2} - E_{c.m.s} 
=0$~~~~~~~~~~~~~~~~~~~~~~~~~~~~~~~~~~~~~~~~(A2.4)
%f_4 = \sqrt{p_{t1}^2 - \mu^2} + \sqrt{p_{t2}^2 - \mu^2} + p_{ee} - E_{c.m.s} 
%=0$(A2.4)
\vskip1mm
$
f_5 = \sqrt{p_{t1}^2 - \mu^2}/p_{t1} - n_1 \cos R_{t1}
=0$~~~~~~~~~~~~~~~~~~~~~~~~~~~~~~~~~~~~~~~~~~~~~~~~~~~~~~~~~~~~~~~~~~~~~~~~~(A2.5)
\vskip0mm
$
f_6 = \sqrt{p_{t2}^2 - \mu^2}/p_{t2}  - n_2 \cos R_{t2}
=0$~~~~~~~~~~~~~~~~~~~~~~~~~~~~~~~~~~~~~~~~~~~~~~~~~~~~~~~~~~~~~~~~~~~~~~~~~~(A2.6)
\vskip1mm

\subsection*{Topology 3, mass fit only}
~~~~~~Definitions:\\
$p_{t1}$ - measured tachyon candidate 1 track momentum;\\
$p_{t2}$ - measured tachyon candidate 2 track momentum;\\
$R_{t1}$ - measured radius of the anomalous ring associated with
the tachyon candidate~1;\\
$R_{t2}$ - measured radius of the anomalous ring associated with
the tachyon candidate~2;\\
$n_1$ and $n_2$ - refraction indices of the radiators in which the
corresponding rings were found;\\
$\mu$ - unknown (common) tachyon mass parameter;\\
\vskip1mm
Equations:
\vskip1mm
\hspace{-4.5mm}
$
f_1 = \sqrt{p_{t1}^2 - \mu^2}/p_{t1} - n_1 \cos R_{t1}
=0$~~~~~~~~~~~~~~~~~~~~~~~~~~~~~~~~~~~~~~~~~~~~~~~~~~~~~~~~~~~~~~~~~~~~~~~~~~~~~~~~~~~~(A3.1)
\vskip1mm
\hspace{-4.5mm}
$
f_2 = \sqrt{p_{t2}^2 - \mu^2}/p_{t2} - n_2 \cos R_{t2}
=0$~~~~~~~~~~~~~~~~~~~~~~~~~~~~~~~~~~~~~~~~~~~~~~~~~~~~~~~~~~~~~~~~~~~~~~~~~~~~~~~~~~~~(A3.2)

and a similar equation for each additional Cherenkov ring found.

\newpage
\noindent
{\bf Table 1.} Some input and output parameters of the kinematic fit of events
of topology~1.
\begin{center}
\begin{tabular}{ |c |c |c |c |c |c |c |c |c |c |c|}
\hline
%          &&&&&&&&&&\\
 Run  &    &$p_1$ meas&$p_1$ fit& $\mu$,  &$E_\gamma^{ISR}$, &  fit   & $r_1$ meas  &$r_1$ fit&$dE/dx$, &$dE/dx$,\\
Event &Beam &$p_2$ meas&$p_2$ fit& error  &    error         &$\chi^2$,& $r_2$ meas &$r_2$ fit&   meas  &  fit \\
      &     &         &         &           &                &  ndf   &  (mrad)     & (mrad)  &  (mip)  & (mip) \\
\hline
%      &     &         &         &         &                  &        &             &         &         &      \\
%69535 &80.7 &$50\pm6$ &  49.0   &  2.32   &     51.3         &    2.0 &$78\pm2$     & 78      &   1.85  &  1.67\\
%1530  &     &Not meas.&  2.76   &$\pm0.24$&   $\pm 0.9$      &     4  &$1010\pm24$  & 998     &$\pm0.10$&      \\
      &     &         &         &         &                  &        &             &         &         &       \\
71768 &86.2 &$92\pm19$&  73.8   &  4.29   &                  &  5.0   &  $85 \pm 3$ &   85    &   3.28  &  3.00 \\
 6277 &     &Not meas.&  12.9   &$\pm0.31$&                  &   6    &  $347\pm15$ &  347    &$\pm0.12$&       \\
      &     &         &         &         &                  &        &$741^*\pm19$ & $737^*$ &         &       \\
      &     &         &         &         &                  &        &             &         &         &       \\
%83700 &94.6 &$76\pm16$&  60.4   &  3.92   &     36.4         &  10.9  &$90\pm4$     & 90      &   3.30  &  3.18\\
% 574  &     &Not meas.&  15.6   &$\pm0.25$&   $\pm 2.1$      &   4    &$265\pm10 $  & 262     &$\pm0.14$&      \\
%      &     &         &         &         &                  &        &$700^*\pm18$ & $707^*$ &         &      \\
84451 &94.6 &$64\pm 4$&  67.9   &  4.98   &     19.9         &  13.7  &$102\pm 5 $  &  96     &   Not   &  2.86 \\
 478  &     &$10\pm 1$&  11.1   &$\pm0.27$&   $\pm 3.9$      &   6    &$460\pm 16$  & 470     &  meas.  &       \\
      &     &         &         &         &                  &        &$786^*\pm20$ & $793^*$ &         &       \\
      &     &         &         &         &                  &        &             &         &         &       \\
%85618 &94.6 &$67\pm12$&  81.8   &  4.31   &                  &  6.1   &$ 80  \pm 3$ &  81     &   3.16  &  3.10\\
%21487 &     &Not meas.&  13.2   &$\pm0.35$&                  &   5    &$342  \pm13$ &  338    &$\pm0.12$&      \\
%      &     &         &         &         &                  &        &$722^*\pm19$ & $734^*$ &         &      \\
%      &     &         &         &         &                  &        &             &         &         &      \\
86748 &94.6 &$13\pm1$ &  12.9   &  4.19   &                  &   2.0  &$335 \pm14$  & 337     &   Not   &  3.02 \\
 757  &     &$87\pm10$&  82.1   &$\pm0.28$&                  &    6   &$ 79 \pm 3$  &  80     &  meas.  &       \\
      &     &         &         &         &                  &        &$747^*\pm19$ & 733$^*$ &         &       \\
      &     &         &         &         &                  &        &             &         &         &       \\
%88024 &94.6 &Not meas.&  78.0   &  4.69   &                  &   8.8  &$80\pm3$     & 86      &   3.19  &  3.16 \\
% 1284 &     &$19\pm1$ &  17.0   &$\pm0.23$&                  &   5    &$298\pm11 $  & 286     &$\pm0.11$&       \\
101830&96.1 &$54\pm5$ &  65.1   &  2.75   &                  &  5.1   &$75\pm3$     & 75      &   3.14  &  3.17 \\
15719 &     &Not meas.&  31.1   &$\pm0.27$&                  &   5    &$107 \pm 4$  & 108     &$\pm0.10$&       \\
      &     &         &         &         &                  &        &             &         &         &       \\
102500&96.1 &$59\pm 7$&  78.4   &  4.65   &     16.6         &  17.5  & $85  \pm 3$ &  86     &   2.50  &  2.34 \\
 9258 &     &Not meas.&   6.4   &$\pm0.39$&   $\pm 3.1$      &   5    &$1010^*\pm27$&$1002^*$&$\pm0.10$&        \\
      &     &         &         &         &                  &        &$816\pm34$   & 817     &         &       \\
      &     &         &         &         &                  &        &             &         &         &       \\
%102875&98.1 &Not meas.&  58.8   &  2.24   &     76.4         &  12.7  &$72\pm2$     & 73      &   0.50  &  2.42\\
% 7310 &     &$5 \pm 1$&   3.4   &$\pm0.11$&   $\pm 0.1$      &   5    &$940^*\pm18$ & $937^*$ &$\pm1.27$&      \\
%      &     &         &         &         &                  &        &$720\pm16$   & 720     &         &      \\
106061&100.1&$94 \pm9$&  93.8   &  5.96   &                  &  3.1   &$ 91  \pm 4$ &  89     &  Not    &  2.22 \\
 3293 &     &$7.6\pm1$&   7.7   &$\pm0.36$&                  &   5    &$885  \pm35$ &  894    &  meas.  &       \\
      &     &         &         &         &                  &        &$1052^*\pm23$&$1055^*$ &         &       \\
      &     &         &         &         &                  &        &             &         &         &       \\
107038&100.1&$74\pm7$.&  90.8   &  5.49   &                  &  6.6   &  $87 \pm 4$ &  87     &   3.04  &  2.72 \\
 11748&     &Not meas.&  10.2   &$\pm0.41$&                  &   5    &$583  \pm22$ &  574    &$\pm0.11$&       \\
      &     &         &         &         &                  &        &             &         &         &       \\
116892&103.0&Not meas.&  85.0   &  4.89   &                  &  4.4   &  $83 \pm 3$ &  85     &   2.87  &  3.08 \\
 5928a&     &$21\pm2$ &  19.3   &$\pm0.29$&                  &   7    &  $266\pm12$ &  263    &$\pm0.17$&       \\
      &     &         &         &         &                  &        &$702^*\pm17$ & $707^*$ &         &       \\
      &     &         &         &         &                  &        &             &         &         &       \\
116892&103.0&$140\pm24$&  98.8  &  4.14   &                  &  4.4   & $75  \pm 2$ &  75     &   Not   &  2.31 \\
 5928b&     &Not meas.&   5.6   &$\pm0.50$&                  &   7    &$1021^*\pm27$& $1015^*$&   meas. &       \\
      &     &         &         &         &                  &        & $831\pm 30$ &  837    &         &       \\
%117471&103.0&$94\pm 9$&  95.4   &  4.36   &     4.5          &  5.8   &$77\pm2$     &  77     &   1.55  &  1.65\\
% 1512 &     &Not meas.&   5.1   &$\pm0.54$&   $\pm 7.1$      &   4    &$970\pm18 $  &1021     &$\pm0.08$&      \\
      &     &         &         &         &                  &        &             &         &         &       \\
%117637&103.0&$ 7\pm1$ &   7.1   &  1.53   &                  &   4.3  &$225\pm 7$   & 226     &   3.43  &  3.21\\
% 4545 &     &Not meas.&  96.0   &$\pm0.27$&                  &    5   &$ 69 \pm 2$  &  64     &$\pm0.12$&      \\
\hline
\end{tabular}
\end{center}
* Asterisks designate liquid radiator rings.
\newpage
Remarks to Table~1 (valid also for subsequent tables):
\begin{itemize}
\item [1).] Beam and tachyon momenta $p_1, p_2$ in tables are given
in GeV/$c$, tachyon mass parameters are given in GeV/$c^2$,
the ISR photon energies are given in GeV.
 \item [2).] If the ISR photon energy is absent in the corresponding column,
this means that it was calculated to be below 1.5 GeV and it is
dropped from the fit.
 \item [3).] If 3rd (4th) anomalous ring was found in a given event, its
parameters are quoted in the 3rd (4th) line of the event raw.
 \item [4).] The number of constraint equations equals to 4 plus the number
of the anomalous rings found. In events with the (common) track ionization
measured an additional constraint, based on the ionization, is added.
 \item [5).] The number of unknowns equals to 1 (tachyon mass parameter
common for both tachyons)
plus number of unmeasured tachyon momenta (1 in most of the events of
topology~1) plus 1 if an ISR photon momentum (energy) participates in the fit.
% \item [6).] We give no post-fit errors of the measured parameters since they
%are in most of cases practically equal to the input ones (being
%slightly decreased by the fit).
\end{itemize}

%\newpage
%\subsection*{Topology 2}
\noindent
{\bf Table 2.} Some input and output parameters of the kinematic fit of events
of topology~2.
\begin{center}
\begin{tabular}{ |c |c |c |c |c |c |c |c |c |c |}
\hline
          &&&&&&&&&\\
 Run  &     & $p_1$ meas &$p_1$ fit& $\mu$,   &   Final $e^+e^-$     &  Mass  &  fit    & $r_1$ meas&$r_1$ fit\\
Event &Beam & $p_2$ meas &$p_2$ fit& error   &longitudinal $\Delta p$,&  of   &$\chi^2$,& $r_2$ meas&$r_2$ fit\\
      &     &            &         &          &      error           &$e^+e^-$&  ndf    &  (mrad)   & (mrad)  \\
\hline
%      &     &            &         &          &                      &        &         &           &         \\
%76039 &91.6 &$7.0\pm0.1$ &  7.14   &  0.306   &          8.3         & 169.7  &  5.7    & $81\pm4$  &  75     \\
% 435  &     &$6.2\pm0.1$ &  6.16   &$\pm0.026$&       $\pm 0.1$      &$\pm0.2$&  3      & $77\pm3$  &  80     \\
      &     &            &         &          &                      &        &         &           &         \\
77190 &91.6 &$5.1\pm0.1$ &  5.03   &  0.296   &          1.2         &  173.0 &  5.3    & $84\pm3$  &  86     \\
 2950 &     &$4.7\pm0.3$ &  5.20   &$\pm0.019$&       $\pm 0.1$      &$\pm0.2$&  3      & $86\pm4$  &  84     \\
      &     &            &         &          &                      &        &         &           &         \\
%80957 &91.6 &$ 3.6\pm0.1$&  3.57    &  0.165   &         1.5         &  175.7 &  2.0    & $80\pm3$  &  77     \\
%10048 &     &$ 3.9\pm0.1$&  3.94   &$\pm0.013$&       $\pm 0.1$      &$\pm0.2$&  3      & $72\pm3$  &  75     \\
%      &     &            &         &          &                      &        &         &           &         \\
%83450 &94.6 &$92.2\pm14.4$& 93.1   &   5.92   &          1.9         &        &  2.1    &$ 89\pm4$  &  89     \\
% 1745 &     &$86.9\pm 5.7$& 94.3   &$\pm 0.33$&       $\pm 0.5$      &        &  4      &$ 88\pm4$  &  88     \\
%      &     &            &         &          &                      &        &         &           &         \\
85371 &94.6 &$9.6\pm0.2$ &  9.48   &  0.881   &          1.0         & 170.1  &  4.8    &$118\pm5$  &  112    \\
 626  &     &$9.6\pm0.2$ &  9.71   &$\pm0.039$&       $\pm 0.1$      &$\pm0.3$&  3      &$104\pm4$  &  110    \\
      &     &            &         &          &                      &        &         &           &         \\
88975 & 94.6&$4.4\pm0.1$ &  4.41   &  0.155   &          1.1         & 180.7  &  0.4    & $72\pm3$  &  71     \\
  84  &     &$4.1\pm0.1$ &  4.09   &$\pm0.015$&       $\pm 0.1$      &$\pm0.1$&  3      & $72\pm3$  &  73     \\
      &     &            &         &          &                      &        &         &           &         \\
105033&100.1&$6.7\pm0.1$ &  6.61   &  0.284   &          1.4         & 187.6  &  7.3    & $76\pm2$  &  74     \\
 3999 &     &$5.8\pm0.1$ &  5.97   &$\pm0.023$&       $\pm 0.1$      &$\pm0.2$&  3      & $78\pm4$  &  78     \\
      &     &            &         &          &                      &        &         &           &         \\
110546&102.5&$100.5\pm6.5$& 102.4  &  4.30    &                      &        &  4.7    & $75\pm3$  &  75     \\
 15356&     &$103.3\pm6.7$& 102.5  &$\pm0.46 $&                      &        &  5      & $75\pm4$  &  75     \\
      &     &            &         &          &                      &        &         &           &         \\
115187&103.0&$  6.0\pm0.1$&  5.95  &  0.268   &         1.6          & 194.2  &  7.2    & $75\pm3$  &  77     \\
 4860 &     &$  5.7\pm0.1$&  5.82  &$\pm0.022$&      $\pm 0.1$       &$\pm0.2$&  3      & $78\pm4$  &  77     \\
      &     &            &         &          &                      &        &         &           &         \\
\hline
\end{tabular}
\end{center}
\newpage
\noindent
{\bf Table 3.} Some input and output parameters of the kinematic fit of events
of topology~3.
\begin{center}
\begin{tabular}{ |c |c |c |c |c |c |c |c |c |}
\hline
      &     &             &              &         &      &          &             &          \\
%          && && && &&\\
 Run   &Beam & $p_e$       &$p_1$ meas    &$p_1$ fit & $\mu$,&    fit   & $r_1$ meas &$r_1$ fit\\
Event  &     &  fit        &$p_2$ meas    &$p_2$ fit & error & $\chi^2$,& $r_2$ meas &$r_2$ fit\\
       &     &             &              &         &           & ndf   &  (mrad)    &  (mrad) \\
\hline
       &     &             &              &       &          &          &            &         \\
41633  & 46.5&    46.5     &$12.03\pm1.51$&  9.62 &   0.302  &    7.9   & $ 72 \pm 3$&    70   \\
 1568  &     &             &$ 4.72\pm0.69$&  4.70 &$\pm0.039$&     6    & $ 89 \pm 5$&    89   \\
       &     &             &              &       &          &          &            &         \\
42495  & 45.6&    45.6     &$21.28\pm2.61$& 15.83 &   0.781  &    6.0   & $ 82 \pm 4$&    79   \\
24853  &     &             &$ 8.89\pm0.71$&  8.78 &$\pm0.073$&     3    & $107 \pm 5$&   109   \\
       &     &             &              &       &          &          &            &         \\
46877  & 45.6&$\approx41.7$&$ 6.46\pm0.31$&  6.43 &   4.71   &    1.1   & $815 \pm32$&   826   \\
 4278  &     &             &$28.99\pm0.76$& 28.70 &$\pm 0.13$&     2    & $181 \pm 7$&   176   \\
       &     &             &              &       &          &          &$987^*\pm26$&$1008^*$ \\
       &     &             &              &       &          &          &            &         \\
49062  & 45.6&$\approx30.4$&$19.69\pm4.02$& 16.33 &   0.896  &    1.2   & $ 84 \pm 4$&    83   \\
 2354  &     &             &$ 3.79\pm0.12$&  4.25 &$\pm0.137$&     1    & $220 \pm10$&   221   \\
       &     &             &              &       &          &          &            &         \\
49316  & 45.6&$\approx31.1$&$18.00\pm3.30$& 20.27 &   0.933  &    0.9   & $ 76 \pm 4$&    77   \\
 584   &     &             &$ 5.67\pm0.47$&  5.49 &$\pm0.082$&     1    & $183 \pm 9$&   182   \\
       &     &             &              &       &          &          &            &         \\
50971  & 45.6&$\approx37.6$&$ 8.83\pm0.89$&  8.67 &   0.297  &    0.3   & $ 72 \pm 4$&    71   \\
 4331  &     &             &$ 4.46\pm1.02$&  4.87 &$\pm0.040$&     1    & $ 86 \pm 5$&    87   \\
       &     &             &              &       &          &          &            &         \\
58547  & 45.6&    45.7     &$12.17\pm1.30$& 13.11 &   0.993  &    3.8   & $ 99 \pm 5$&    98   \\
23906  &     &             &$10.90\pm1.22$& 12.56 &$\pm0.073$&     3    & $ 99 \pm 5$&   101   \\
       &     &             &              &       &          &          &            &         \\
 62192 & 46.5&$\approx39.3$&$15.45\pm2.95$& 12.43 &   4.57   &    1.7   & $380 \pm16$&   381   \\
 20534 &     &             &$ 7.17\pm0.17$& 7.19  &$\pm 0.18$&     3    & $695 \pm29$&   691   \\
       &     &             &              &       &          &          &$930^{**}\pm28$&$940^{**}$\\
       &     &             &              &       &          &          &$1024^{**}\pm31$&$1059^{**}$\\
       &     &             &              &       &          &          &            &         \\
 79940 & 91.6&    79.5     &$99.6\pm 36.8$& 72.31 &   3.96   &   1.2   & $ 83 \pm 3$&    83   \\
  2890 &     &             &$ 5.36\pm1.65$&  7.35 &$\pm 0.40$&     3    & $580 \pm24$&   571   \\
       &     &             &              &       &          &          &$832^*\pm21$& $847^*$ \\
       &     &             &              &       &          &     & $990^{**}\pm30$&$1007^{**}$\\
       &     &             &              &       &          &          &             &         \\
\hline                                                                                        
\end{tabular}
\end{center}
\newpage
\noindent
{\bf Table 3, Continued}. Some input and output parameters of the kinematic fit of events
of topology~3.
\begin{center}
\begin{tabular}{ |c |c |c |c |c |c |c |c |c |}
\hline
       &     &             &              &         &      &          &             &          \\
 Run   &Beam & $p_e$       &$p_1$ meas    &$p_1$ fit&$\mu,$&   fit    & $r_1$ meas  &$r_1$ fit \\
 Event &     & fit         &$p_2$ meas    &$p_2$ fit& error& $\chi^2$,& $r_2$ meas  &$r_2$ fit\\
       &     &             &              &         &      &   ndf    &    (mrad)   &  (mrad) \\
\hline
       &     &             &              &         &           &       &            &         \\
 104949&100.1&   100.1     &$11.94\pm1.24$& 11.02 &   4.64   &    4.2   & $442 \pm17$&   439   \\
 27827 &     &             &$ 7.35\pm1.89$&  6.79 &$\pm 0.49$&     7    & $742 \pm31$&   756   \\
       &     &             &              &       &          &          &$792^*\pm21$& $778^*$ \\
       &     &             &              &       &          &          &$978^*\pm26$& $961^*$ \\
       &     &             &              &       &          &          &            &         \\
 105892&100.1&   100.1     &$3.67\pm 0.60$&  3.50 & 0.290    &    1.6   &  $104\pm 4$&   104   \\
 13127 &     &             &$3.30\pm 0.23$&  3.48 &$\pm0.033$&     5    &  $104\pm 4$&   104   \\
       &     &             &              &       &          &          &            &         \\
% 114204&103.0&    81.9     &   Not meas.  & 15.24 &  4.68    &    3.5   &$735^*\pm20$& $ 726^*$\\
%%           electron momentum of  64.1 +/- 2.6  from nettfitnw.lis  was used!
%  2668 &     &             &   Not meas.  &  6.21 &$\pm 1.95$&     3    &$1015^*\pm27$&$1029^*$\\
%       &     &             &              &       &          &          & $318\pm12$ &  318    \\
%       &     &             &              &       &          &          & $875\pm35$ &  856    \\
\hline
\end{tabular}
\end{center}

\noindent * Asterisks designate the liquid radiator rings.

\noindent ** Double asterisks designate the quartz radiator rings.

\vskip0.6cm

\newpage
%\vskip-0.5cm
\begin{figure}
\begin{center}
\epsfig{file=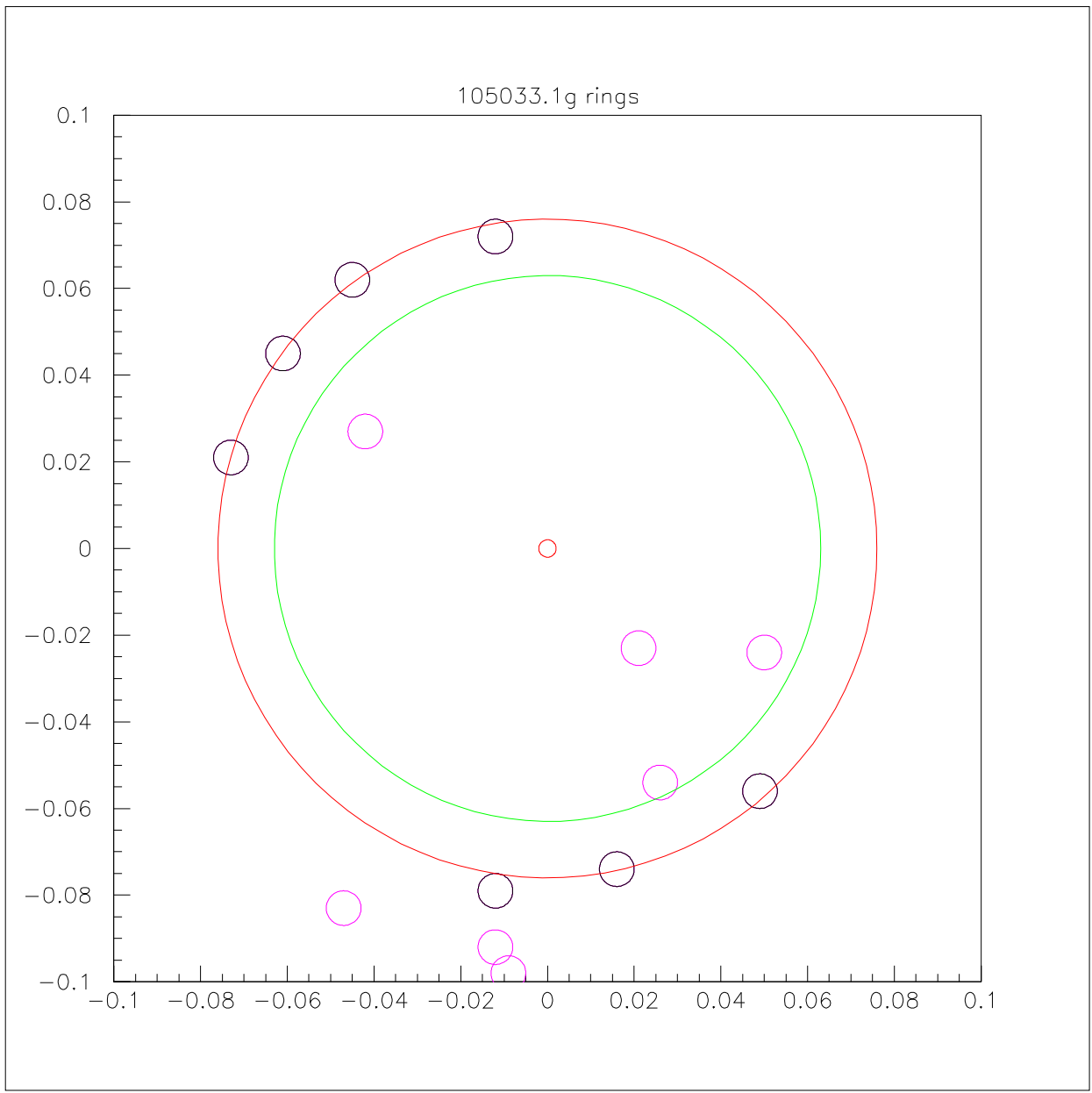,bbllx=20pt,bblly=200pt,bburx=665pt,bbury=570pt,%
width=19cm,angle=0}
\end{center}
\vskip3.2cm
\caption{Gas radiator hit pattern for the 1st track of the event of topology 2b
105033:3999 and a ring of a radius~76 mrad produced by the track. 
Small circles in the hit pattern represent the RICH hits; 
their radii are equal to single photon angular accuracy $\sigma$ 
in the vicinity of the ring under consideration.
The hits pertaining to the anomalous ring are plotted in bold. 
The (inner) green circle marks the position and the size of a standard 
ring (not seen in the pattern). The probability of anomalous ring (marked by 
the red circle) to be fortuitously reconstructed from background hits is below 
$1.2\times10^{-4}$. The small red circle in its center marks the origin of 
the Cherenkov plane, i.e. the position of the image of the track impact point, 
$x=0,~y=0$. The axes in this figure, as well as in all analogous ones below, 
represent coordinates of individual hits on the Cherenkov plane; the axes 
scale values are given in radians.} 
\label{fig:3}
\end{figure}

\newpage
\begin{figure}
\vskip-0.5cm
\begin{center}
\epsfig{file=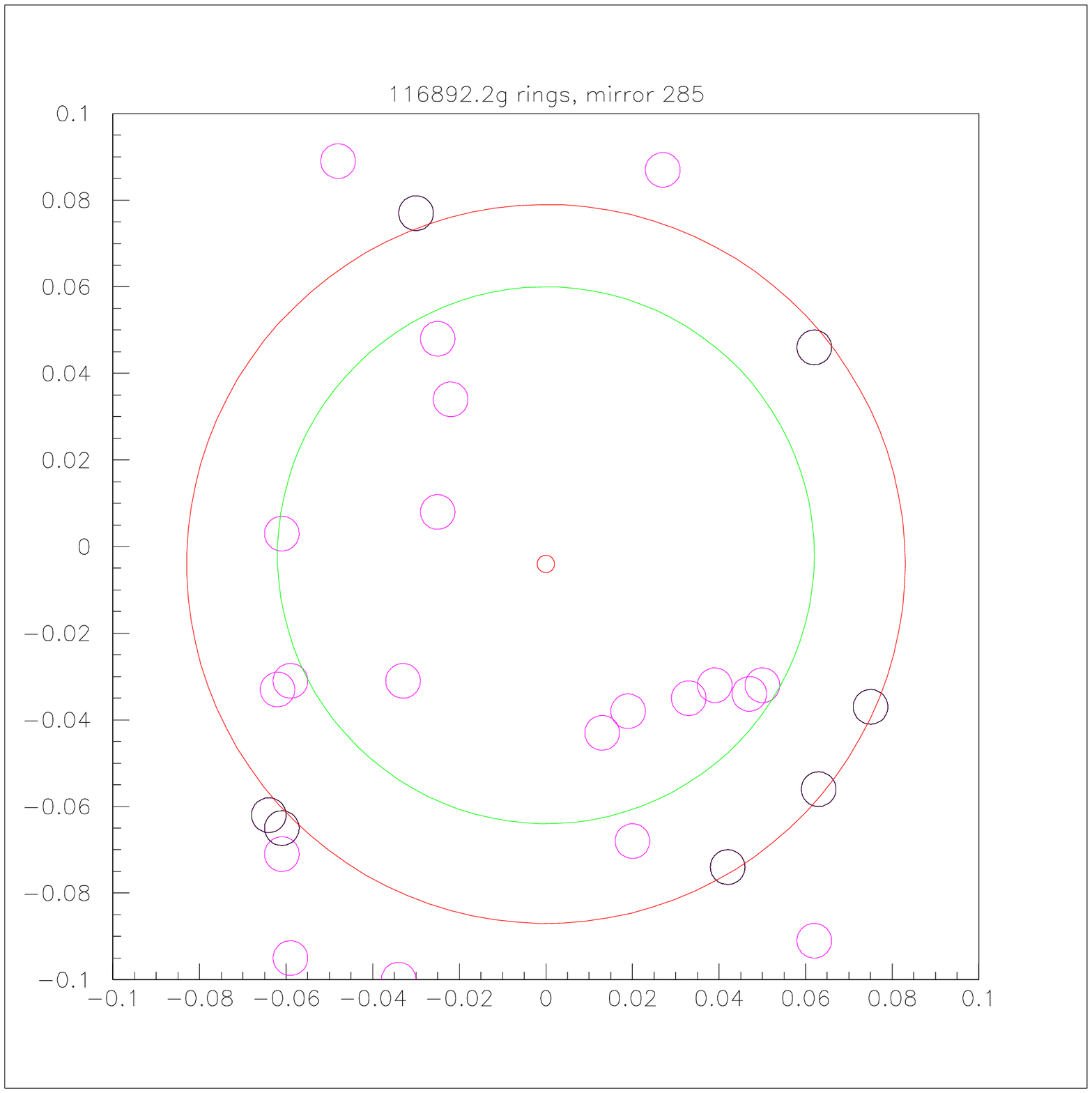,bbllx=20pt,bblly=200pt,bburx=665pt,bbury=570pt,%
width=19cm,angle=0}
\end{center}
\vskip3.2cm
\caption{Gas radiator hit pattern for a track of the event of topology 1a 
116892:5928a, and a ring of a radius 83~mrad produced by the track.
The green circle marks the position and the size of a standard ring (not seen 
in the pattern). The probability of anomalous ring (marked by the red circle) 
to be fortuitously reconstructed from background hits is below 
$9.0\times10^{-3}$. The units in the Cherenkov plane are given in radians, 
the small red circle in its center marks the position of the image of
the track impact point.}
\label{fig:0}
\end{figure}    
\newpage
\begin{figure}
\vskip-0.5cm
\begin{center}
\epsfig{file=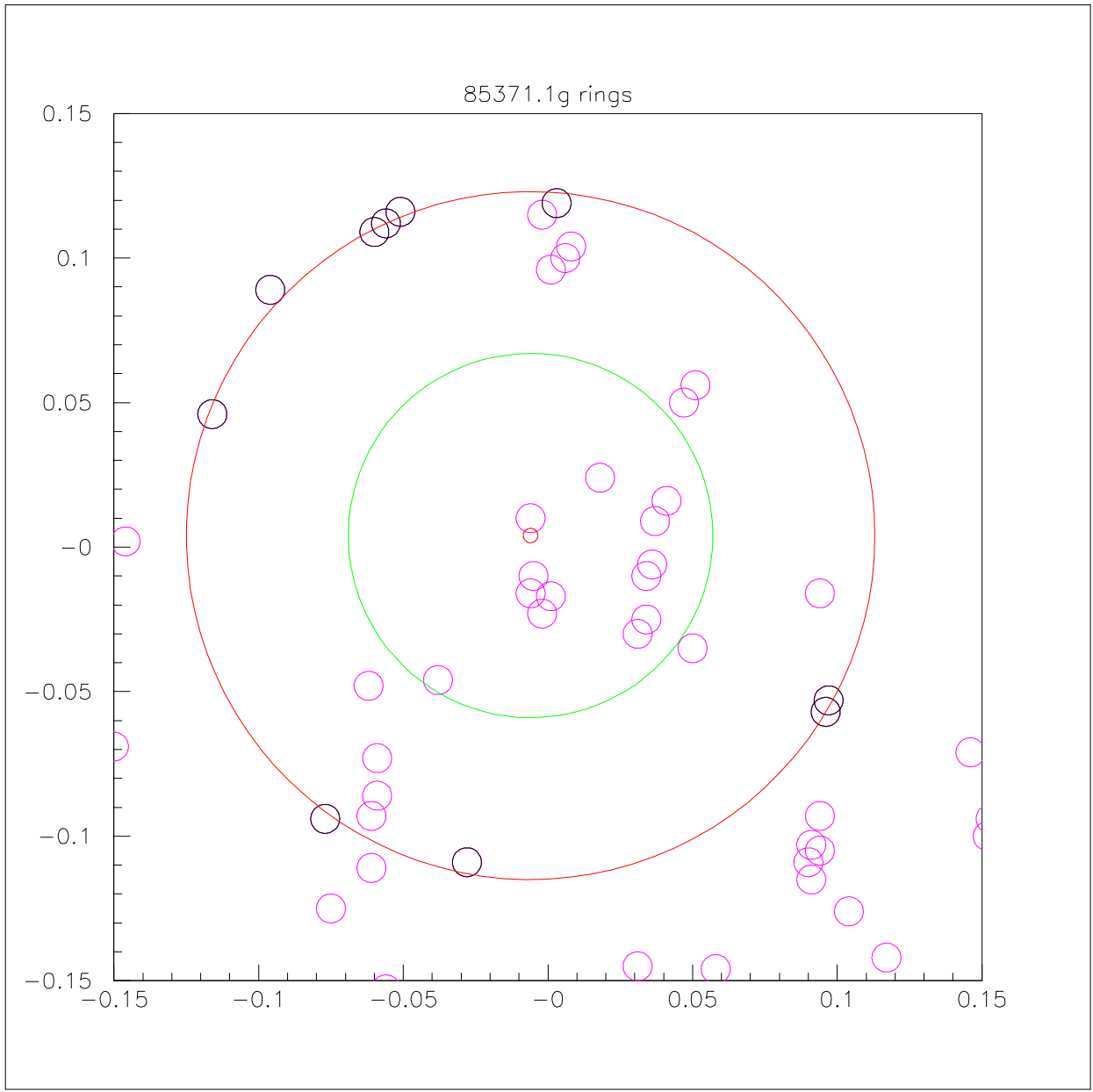,bbllx=20pt,bblly=200pt,bburx=665pt,bbury=570pt,%
width=19cm,angle=0}
\end{center}
\vskip3.2cm
\caption{Gas radiator hit pattern for the 1st track of the event of topology 2b
85371:626, and a ring of a radius 118~mrad produced by the track.
The green circle marks the position and the size of a standard ring 
(not seen in the pattern). The probability of anomalous ring (marked by the red
circle) to be fortuitously reconstructed from background hits is below 0.0158.
The units in the Cherenkov plane are given in radians, the small red circle in
its center marks the position of the image of the track impact point.}
\label{fig:4}
\end{figure}

\newpage
\begin{figure}
\vskip-0.5cm
\begin{center}
\epsfig{file=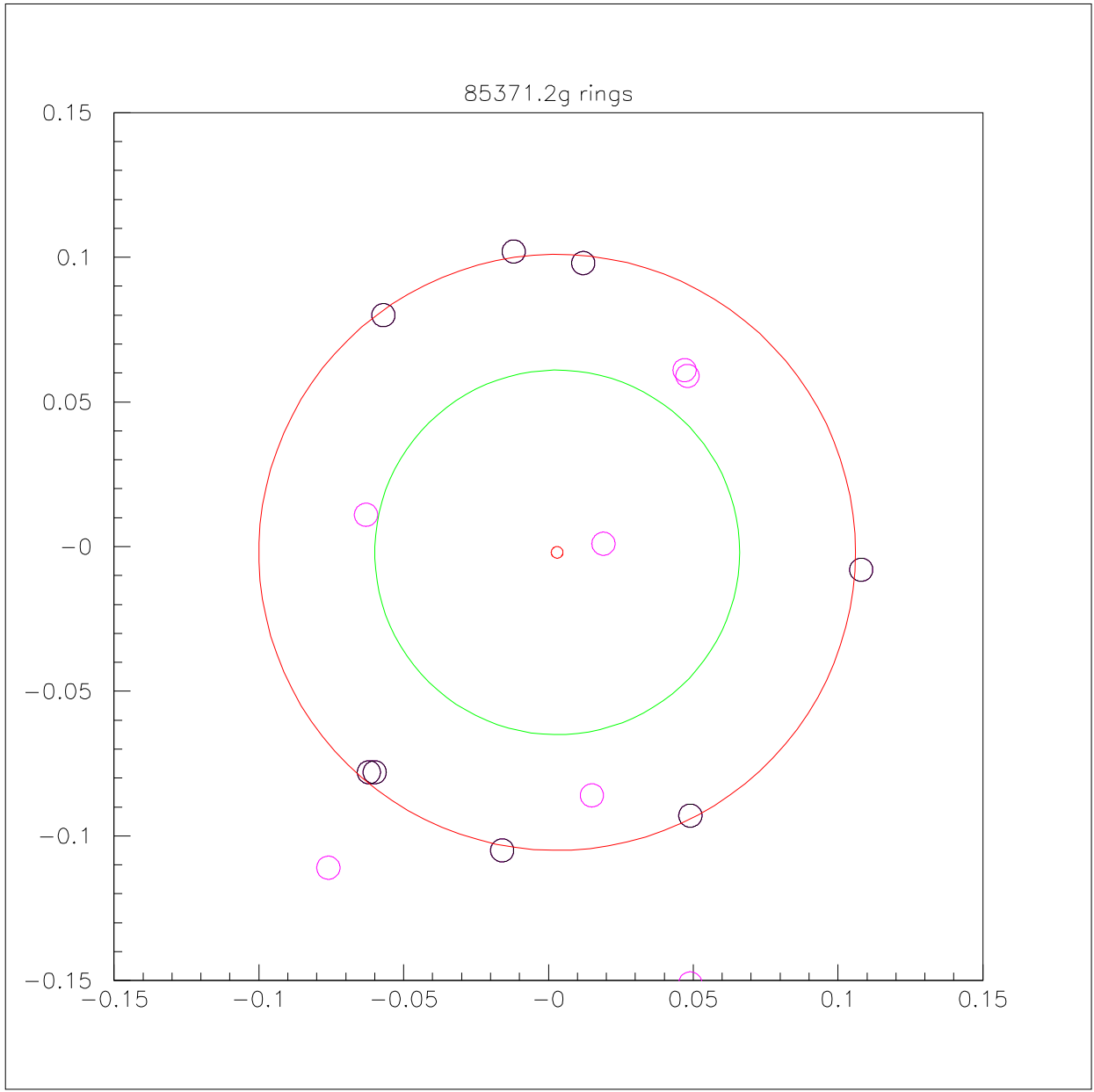,bbllx=20pt,bblly=200pt,bburx=665pt,bbury=570pt,%
width=19cm,angle=0}
\end{center}
\vskip3.2cm
\caption{Gas radiator hit pattern for the 2nd track of the event 85371:626
(the same event as presented by another ring in the previous figure, 
Fig.~\ref{fig:4}), and a ring of a radius 104~mrad produced by this track.
The green circle marks the position and the size of a standard ring (not seen 
in the pattern). The probability of anomalous ring (marked by the red circle) 
to be fortuitously reconstructed from background hits is below 
$6.4\times10^{-5}$. The units in the Cherenkov plane are given in radians, 
the small red circle in its center marks the position of the image of the track 
impact point.}
\label{fig:5}
\end{figure}    

\newpage
\begin{figure}
\vskip 0.5cm
\begin{center}
\epsfig{file=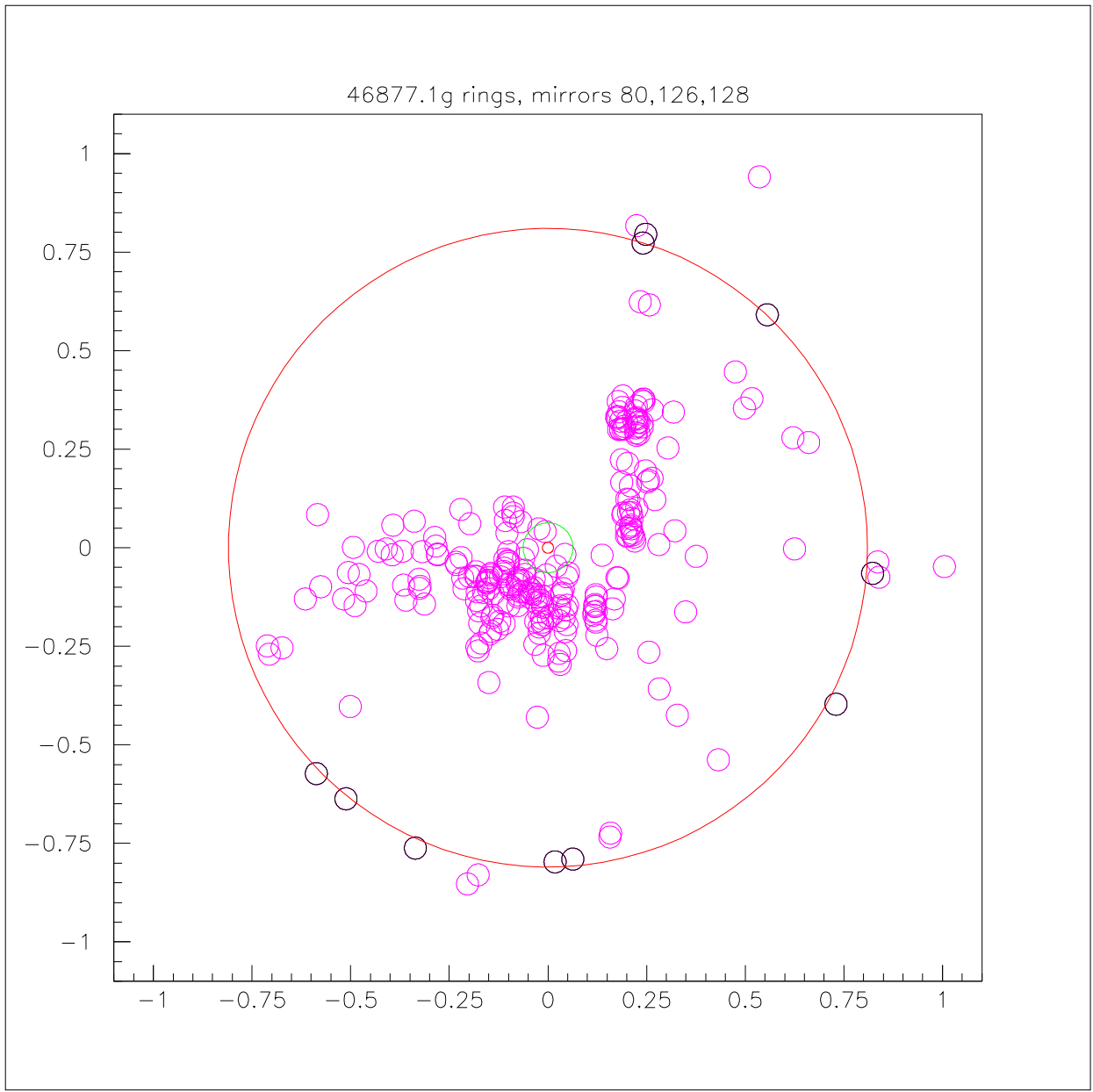,bbllx=20pt,bblly=200pt,bburx=665pt,bbury=570pt,%
width=19cm,angle=0}
\end{center}
\vskip5.0cm
\caption{Gas radiator hit pattern for one of the tracks in a 3-particle jet 
 of an event of topology~3 46877:4278, and a ring of a radius 815~mrad
produced by this track. The green circle in the center of the hit pattern
marks the position and the size of a standard ring
(not seen in the pattern). The probability of anomalous ring (marked by the red
circle) to be fortuitously reconstructed from background hits is below 
$1.7 \times 10^{-4}$.
The units in the Cherenkov plane are given in radians, the small red circle in
its center marks the position of the image of the track impact point.}
  
\label{fig:100}
\end{figure}

\newpage
\begin{figure}
\vskip 0.5cm
\begin{center}
\epsfig{file=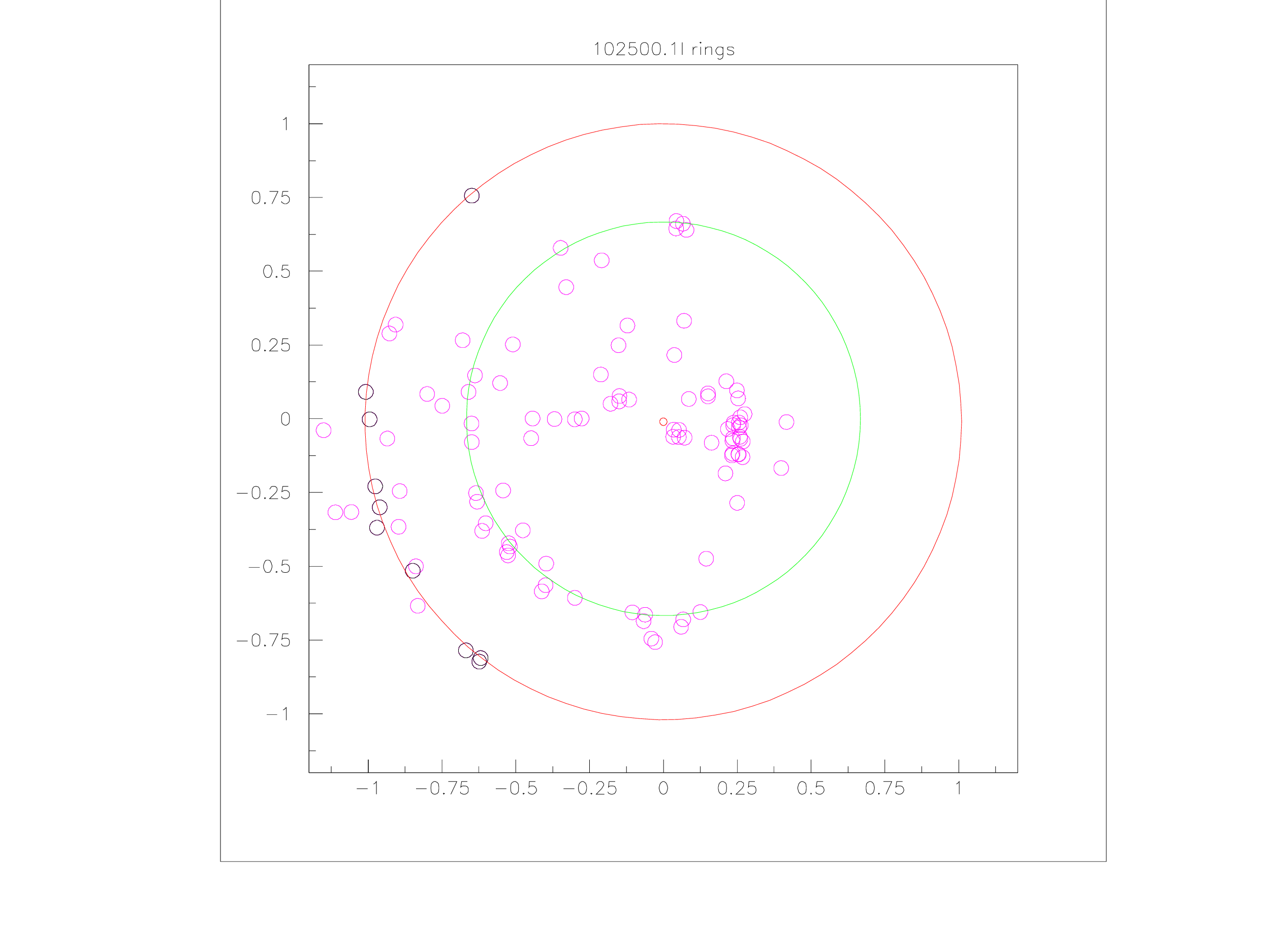,bbllx=90pt,bblly=200pt,bburx=750pt,bbury=420pt,%
width=19cm,angle=0}
\end{center}
\vskip5.0cm
\caption{Liquid radiator hit pattern for a double track of an event of 
topology 1a 102500:9258. Together with an anomalous ring
(indeed, an arc due to the effect of the total inner reflection), having 
radius of 1010~mrad and marked by a red circle, a ring of a radius of 674~mrad
(close to the radius of a standard ring 667~mrad and also presented as an arc) 
marked by a green circle, is seen. The probability of the anomalous ring 
%marked by the red circle 
to be fortuitously reconstructed from background hits is below 
$2.4 \times 10^{-3}$.}
\label{fig:1}
\end{figure}

%\newpage
%\begin{figure}
%\begin{center}
%\epsfig{file=fig4.eps,bbllx=50pt,bblly=180pt,bburx=550pt,bbury=670pt,%
%width=17cm,angle=0}
%\end{center}
%\vskip1.5cm
%\caption{ The principle of the DELPHI Barrel RICH operation.}
%\label{fig:6}
%\end{figure}
\newpage
\begin{figure}
\begin{center}
\epsfig{file=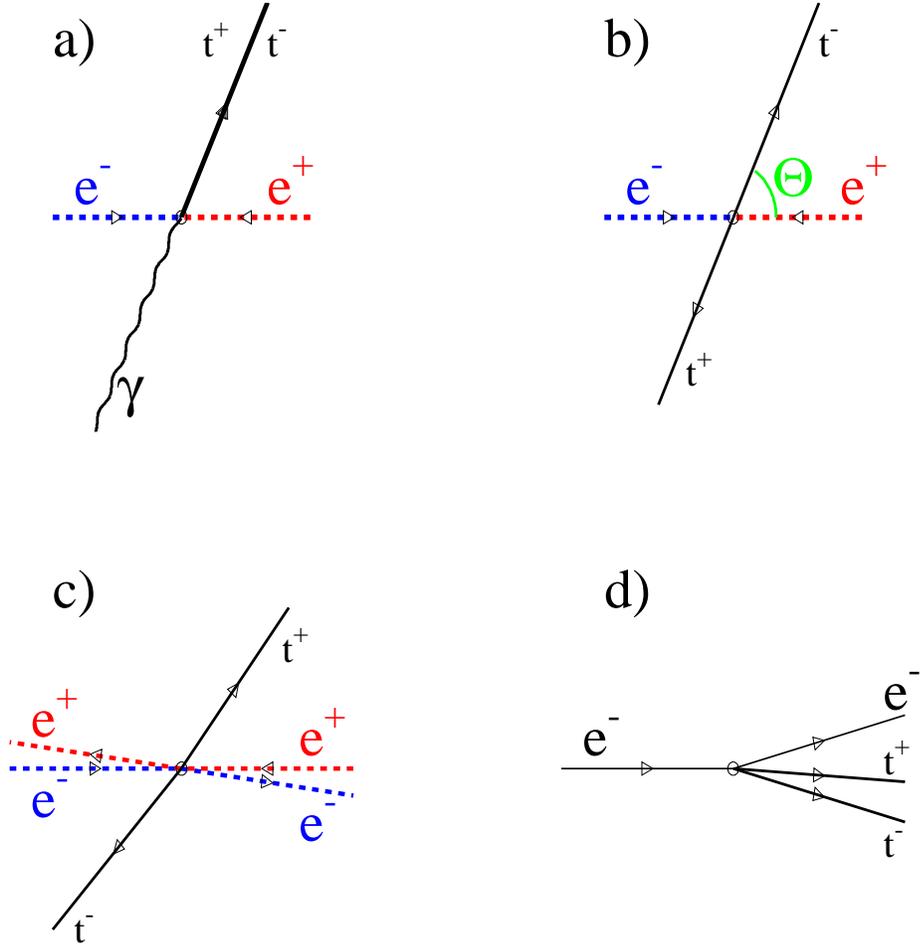,bbllx=0pt,bblly=200pt,bburx=650pt,bbury=570pt,%
width=19cm,angle=0}
\end{center}
%\vskip-1cm
\caption{Reaction diagrams of topologies studied: {\bf a)}~topology~1, 
production of a tachyon-antitachyon pair with nearly-zero opening angle; 
{\bf b)}~topology~2a, back-to-back tachyon-antitachyon production; 
{\bf c)}~topology~2b generated by a different mechanism of the 
tachyon-antitachyon pair production, via a $\gamma~\gamma$ interaction; 
{\bf d)}~topology~3, electroproduction of a tachyon-antitachyon pair on nuclei 
in the detector material. The dashed lines mark beam particles and/or the 
particles that go undetected escaping into the beam pipe.}
\label{fig:6}
\end{figure}    

\newpage
\begin{figure}
\begin{center}
\epsfig{file=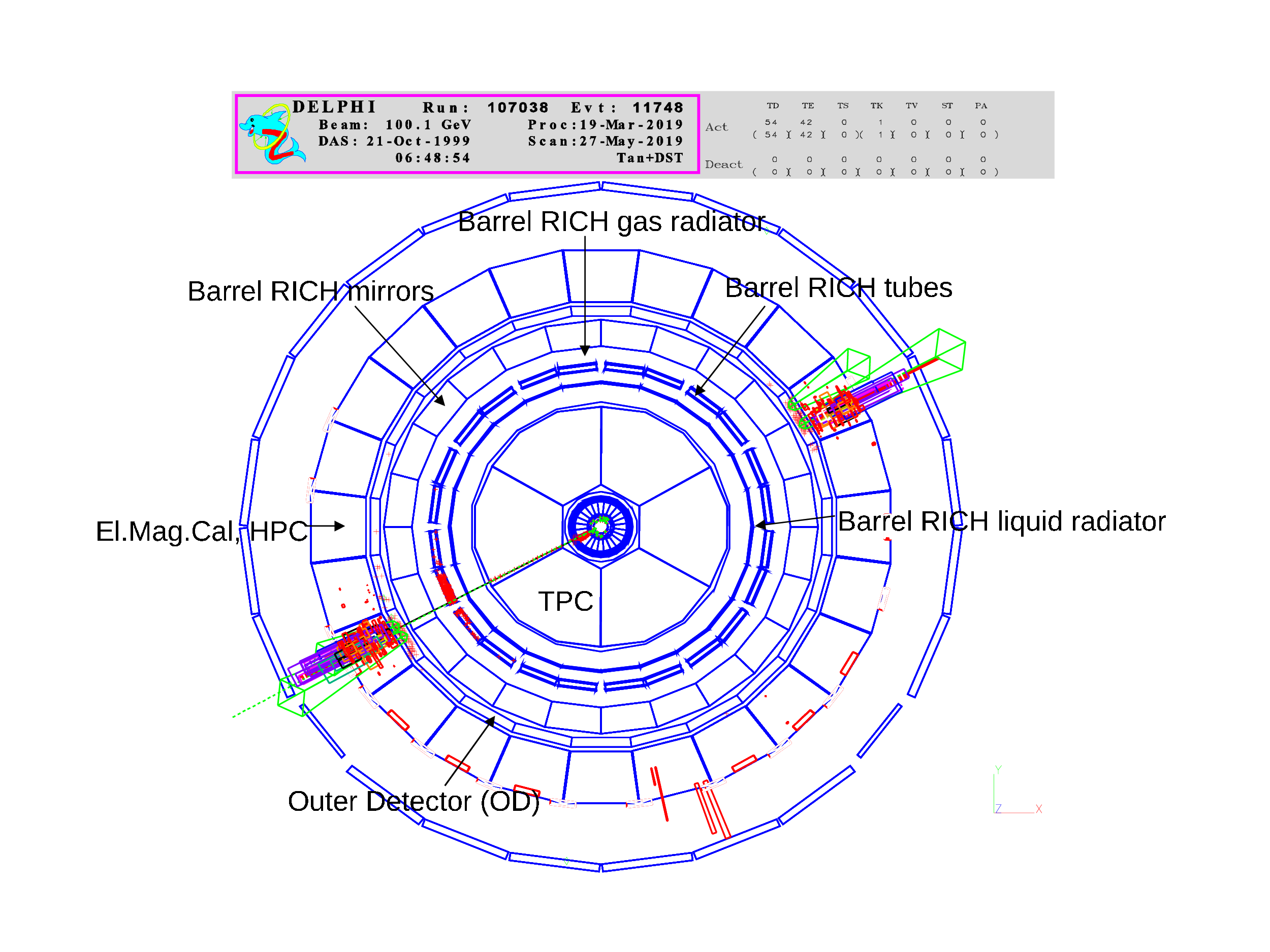,bbllx=100pt,bblly=000pt,bburx=675pt,bbury=400pt,%
width=17cm,angle=0}
\end{center}
\vskip 1cm
\caption{ An example of event of topology 1a of reaction (3.1) with two tracks
non-resolved in the TPC, a general view.
The red hits inside the Barrel RICH drift tubes are Barrel RICH hits (the
tubes will not be shown in next figures).}
\label{fig:61}
\end{figure}

\newpage
\begin{figure}
\begin{center}
\epsfig{file=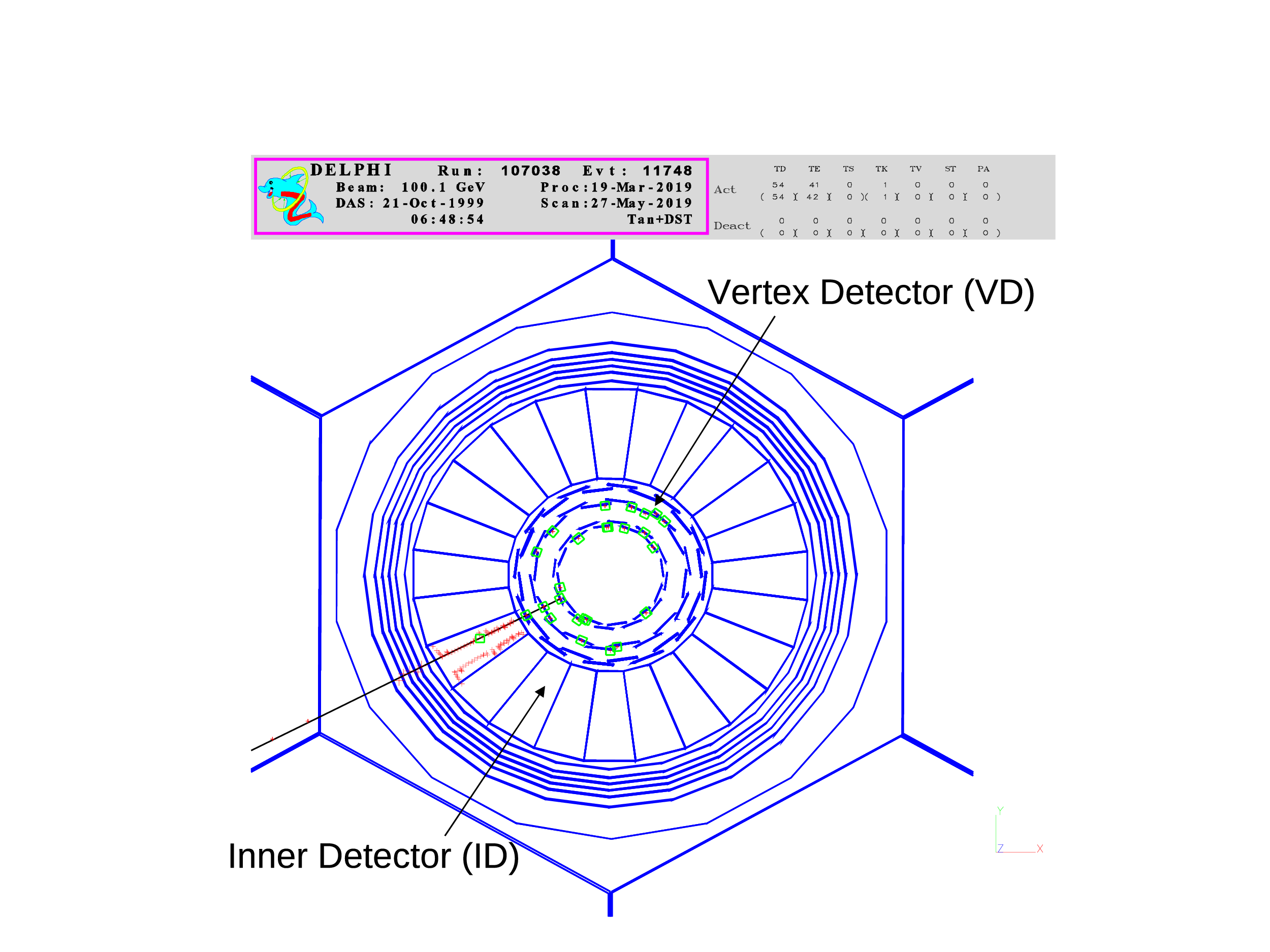,bbllx=100pt,bblly=00pt,bburx=675pt,bbury=500pt,%
width=19cm,angle=0}
\end{center}
\vskip 1cm
\caption{ The same event, the central detector (VD and ID) view.}
\label{fig:161}
\end{figure}

\newpage
\begin{figure}
\begin{center}
\epsfig{file=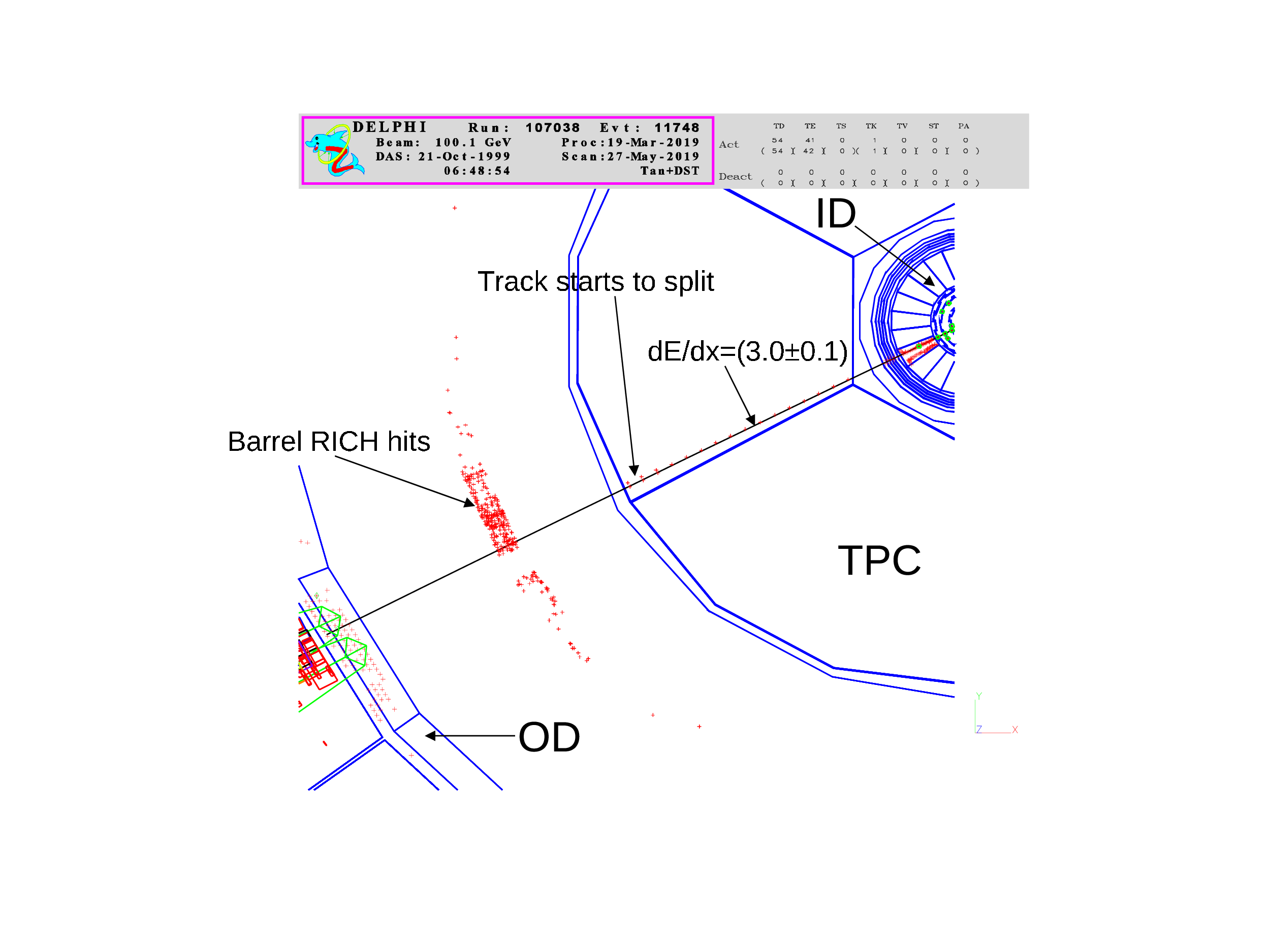,bbllx=100pt,bblly=0pt,bburx=675pt,bbury=500pt,%
width=19cm,angle=0}
\end{center}
%\vskip 3cm
\caption{The same event, the TPC view showing the non-resolved track starting
to split. The $dE/dx$ units are mips (see footnote 7 on page~6).}
\label{fig:261}
\end{figure}

\newpage
\begin{figure}
\begin{center}
\epsfig{file=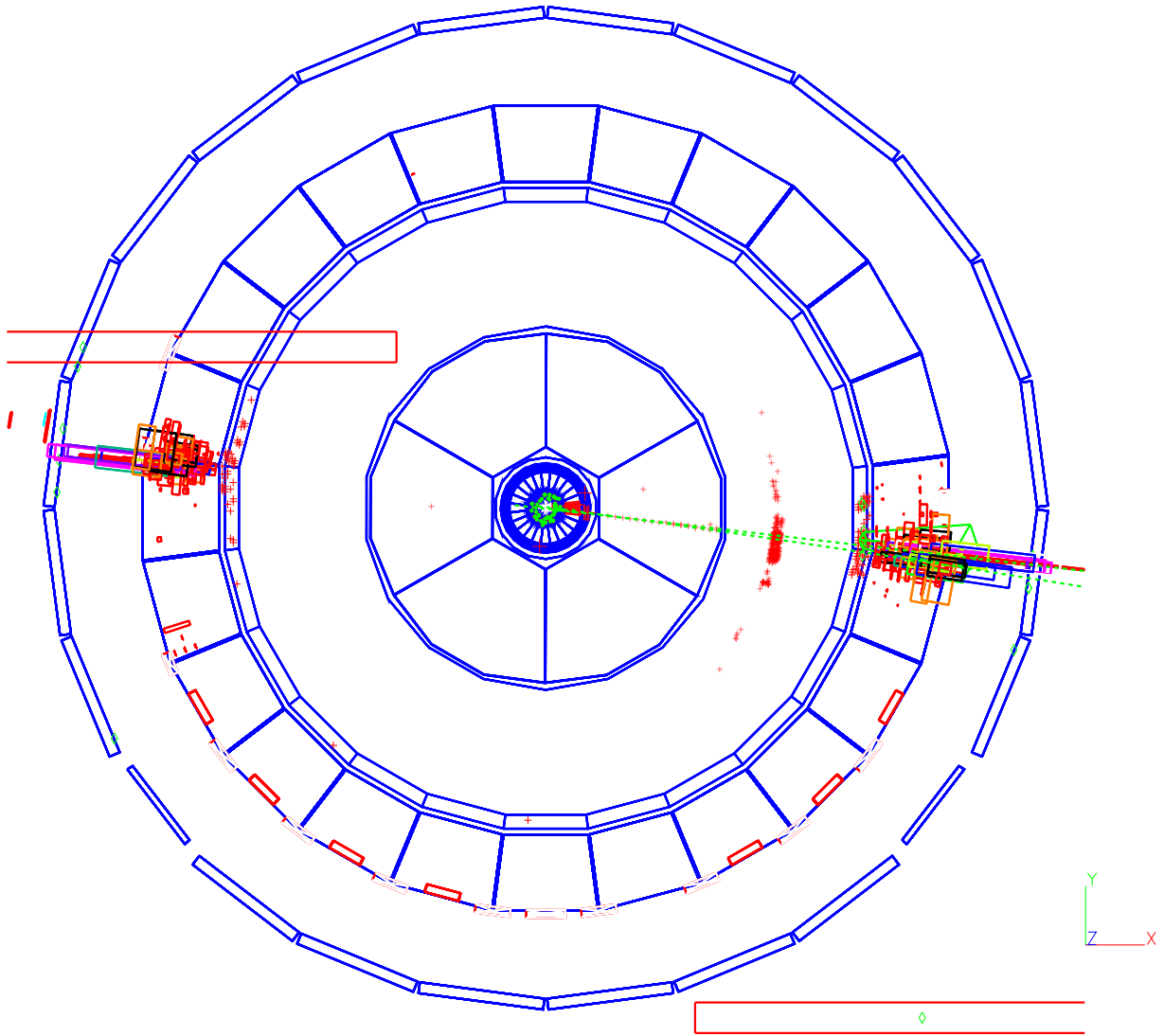,bbllx=0pt,bblly=0pt,bburx=675pt,bbury=700pt,%
width=20cm,angle=0}
\end{center}
\vskip-2cm
\caption{ An example of event of topology 1b of reaction (3.1) with two tracks
resolved in the TPC, a general view.}
\label{fig:61a}
\end{figure}

\newpage
\begin{figure}
\begin{center}
\epsfig{file=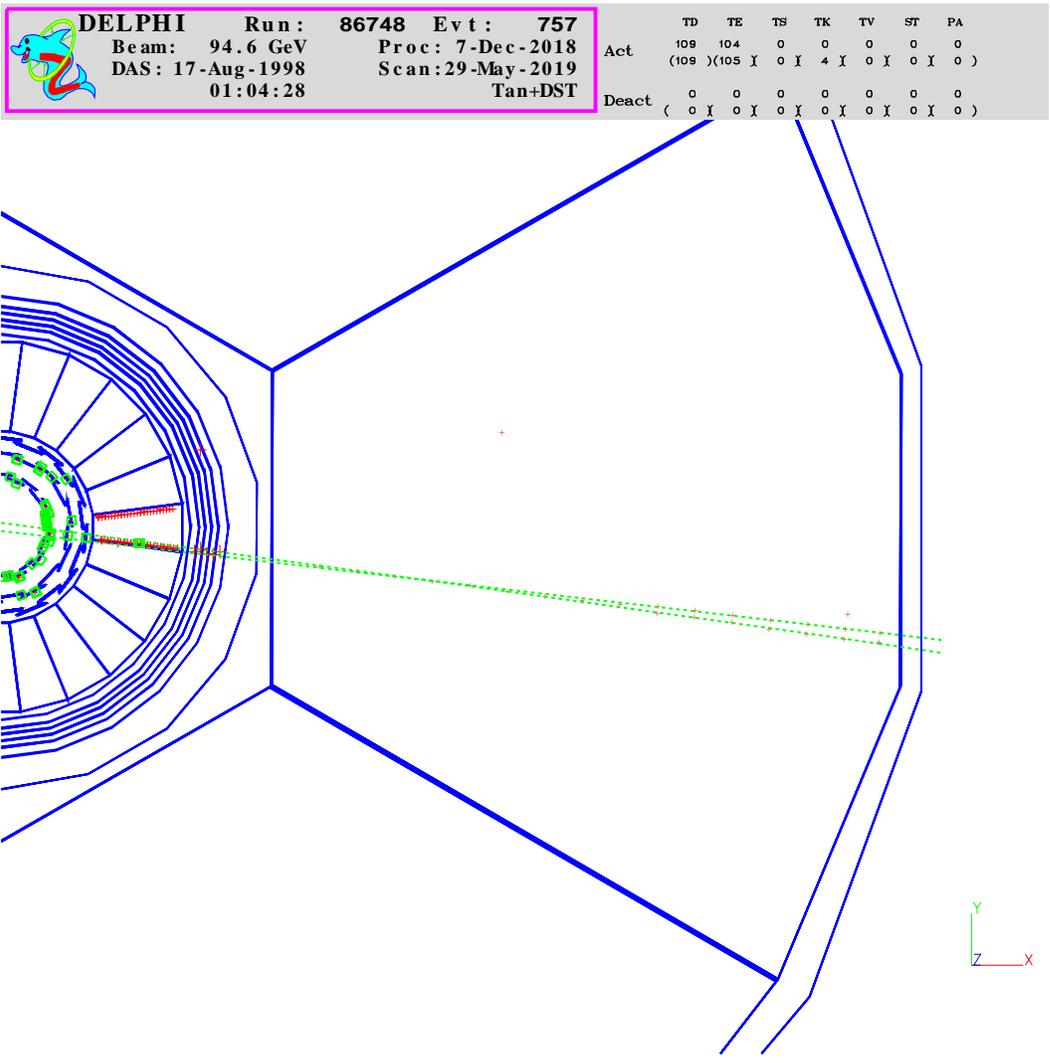,bbllx=0pt,bblly=0pt,bburx=675pt,bbury=500pt,%
width=17cm,angle=0}
\end{center}
\vskip-2cm
\caption{ The same event, the VD, ID and TPC view of the two-particle resolved
jet.}
\label{fig:161a}
\end{figure}

% go to topology 2
\newpage
\begin{figure}
\begin{center}
\epsfig{file=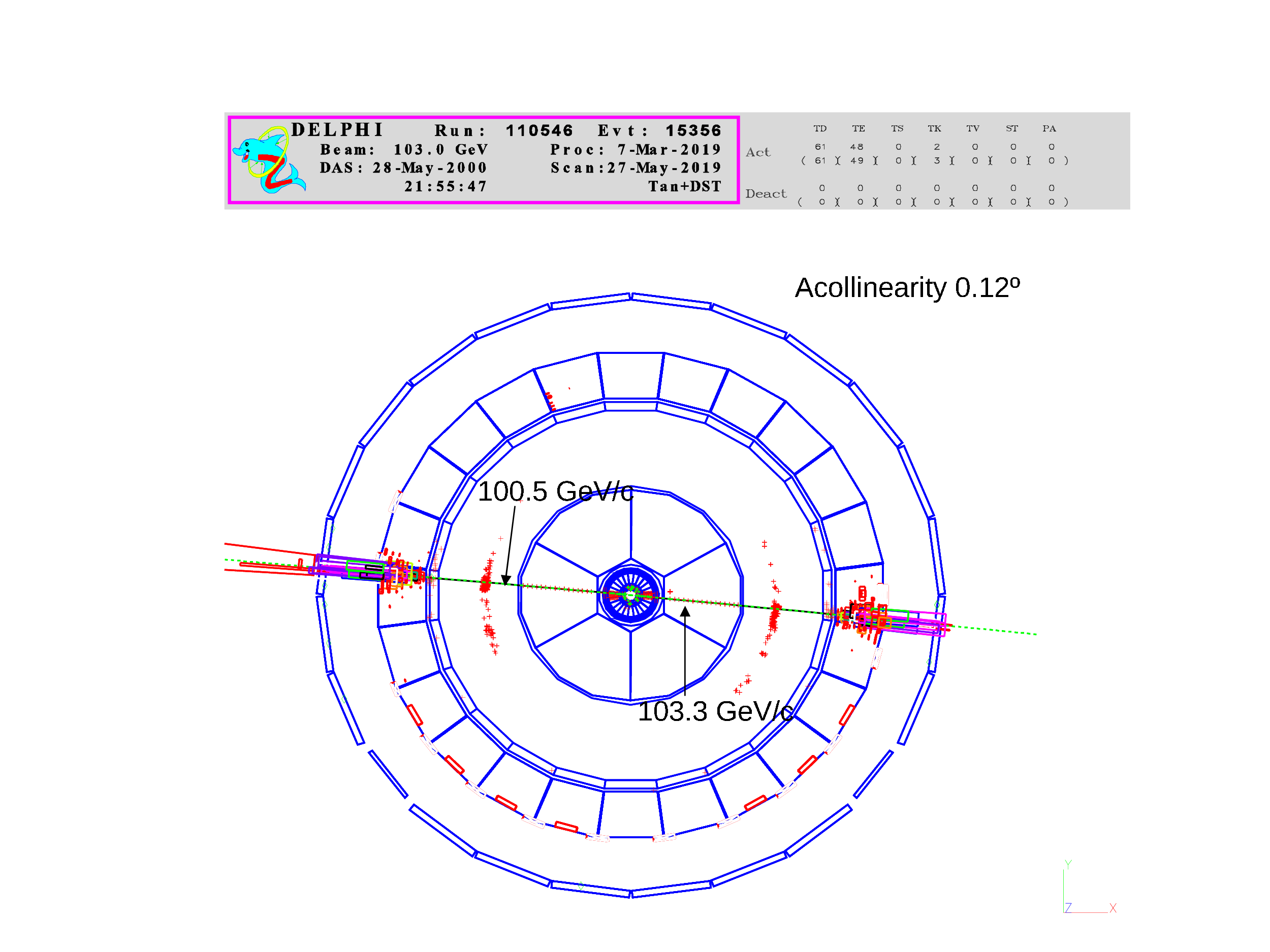,bbllx=120pt,bblly=0pt,bburx=675pt,bbury=500pt,%
width=18cm,angle=0}
\end{center}
\vskip3cm
\caption{ An example of event of topology 2a of reaction 3.2, a general view.}
\label{fig:62a}
\end{figure}

\newpage
\begin{figure}
\begin{center}
\epsfig{file=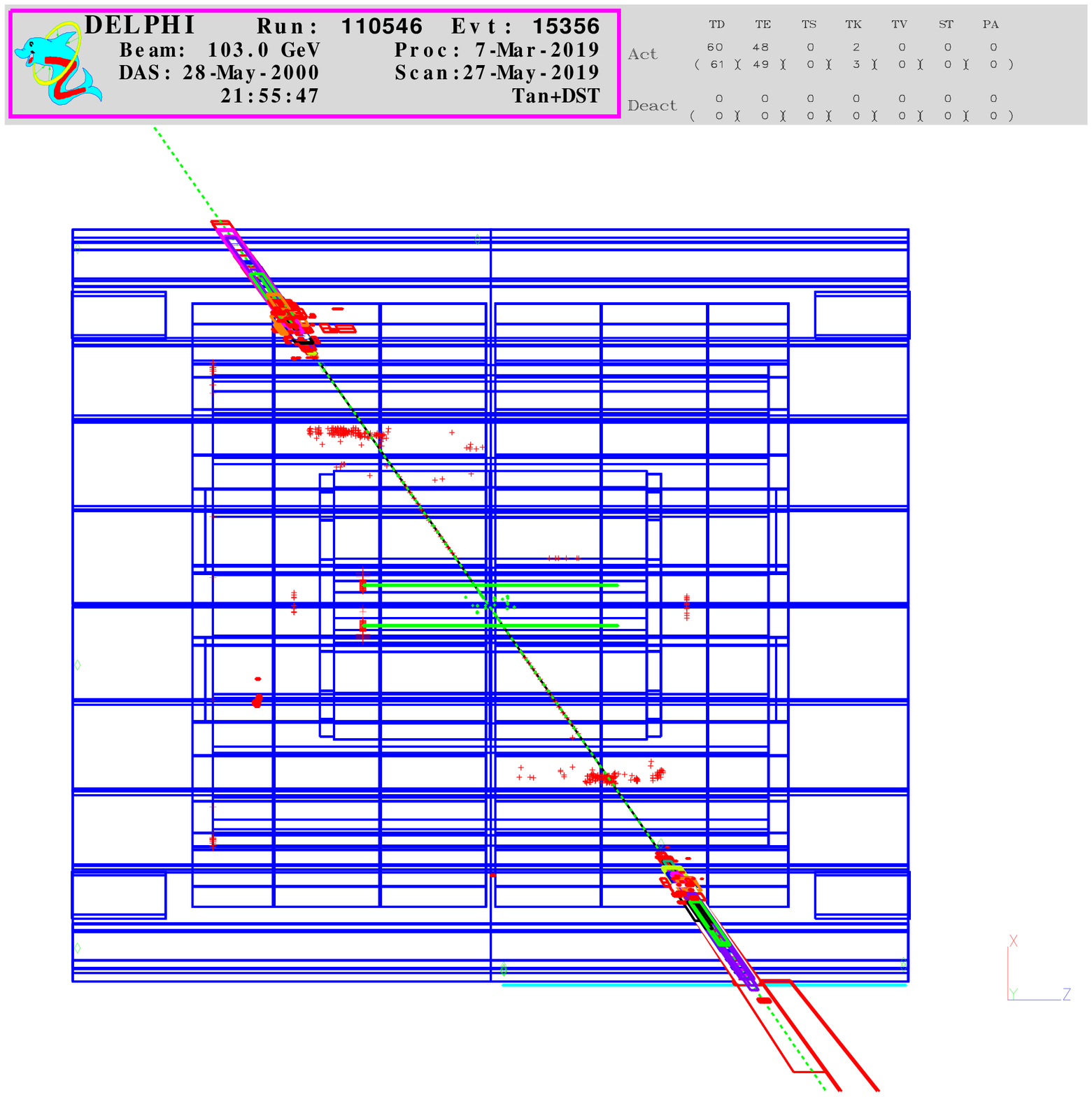,bbllx=0pt,bblly=0pt,bburx=675pt,bbury=800pt,%
width=20cm,angle=0}
\end{center}
\vskip-3cm
\caption{ The same event, the XZ view.}
\label{fig:162a}
\end{figure}

\newpage
\clearpage
\begin{figure}
\begin{center}
\epsfig{file=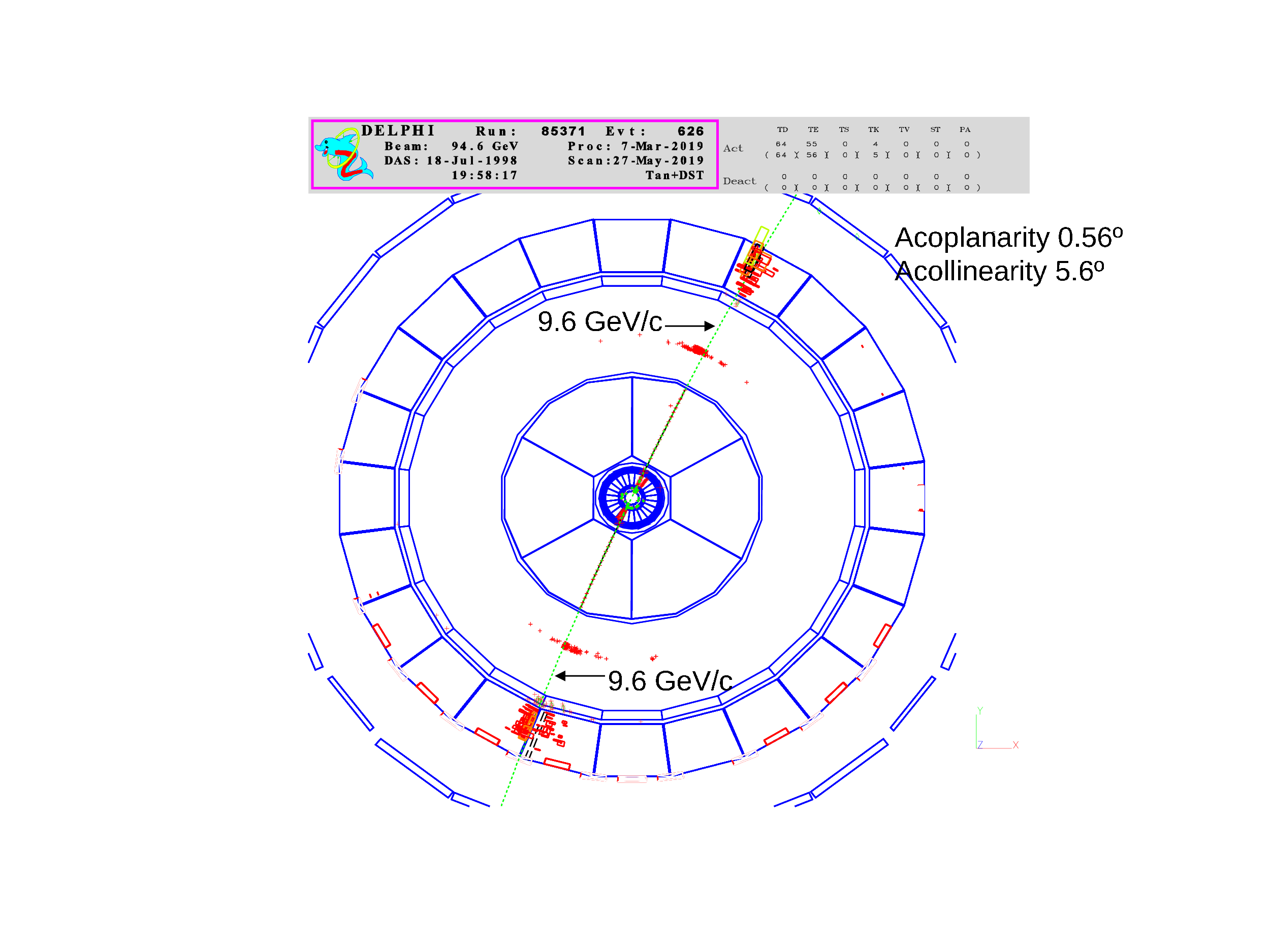,bbllx=120pt,bblly=0pt,bburx=675pt,bbury=500pt,%
width=18cm,angle=0}
\end{center}
\vskip 1cm
\caption{ An example of event of topology 2b of reaction 3.3, a general view.}
\label{fig:62}
\end{figure}

\newpage
\clearpage
\begin{figure}
\begin{center}
\epsfig{file=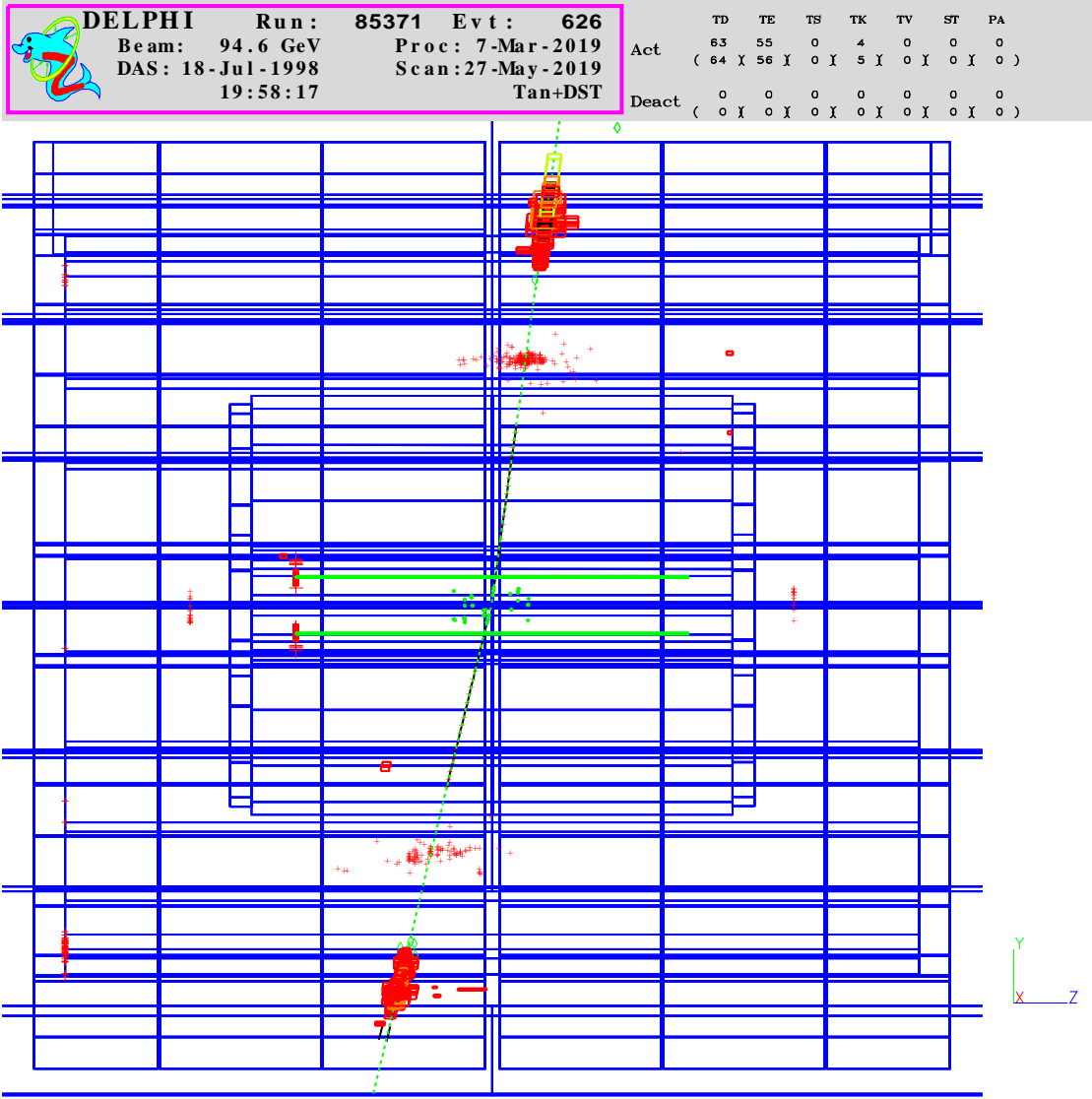,bbllx=0pt,bblly=0pt,bburx=675pt,bbury=700pt,%
width=18cm,angle=0}
\end{center}
\vskip 1cm
\caption{ The same event, the YZ view.}
\label{fig:162}
\end{figure}

%\newpage
%\begin{figure}
%\begin{center}
%\epsfig{file=85371gen.ps,bbllx=150pt,bblly=0pt,bburx=675pt,bbury=570pt,%
%width=20cm,angle=0}
%\end{center}
%\vskip-3cm
%\caption{ Another example of event of topology 2b, a general view.}
%\label{fig:62a}
%\end{figure}

\newpage
\begin{figure}
\begin{center}
\epsfig{file=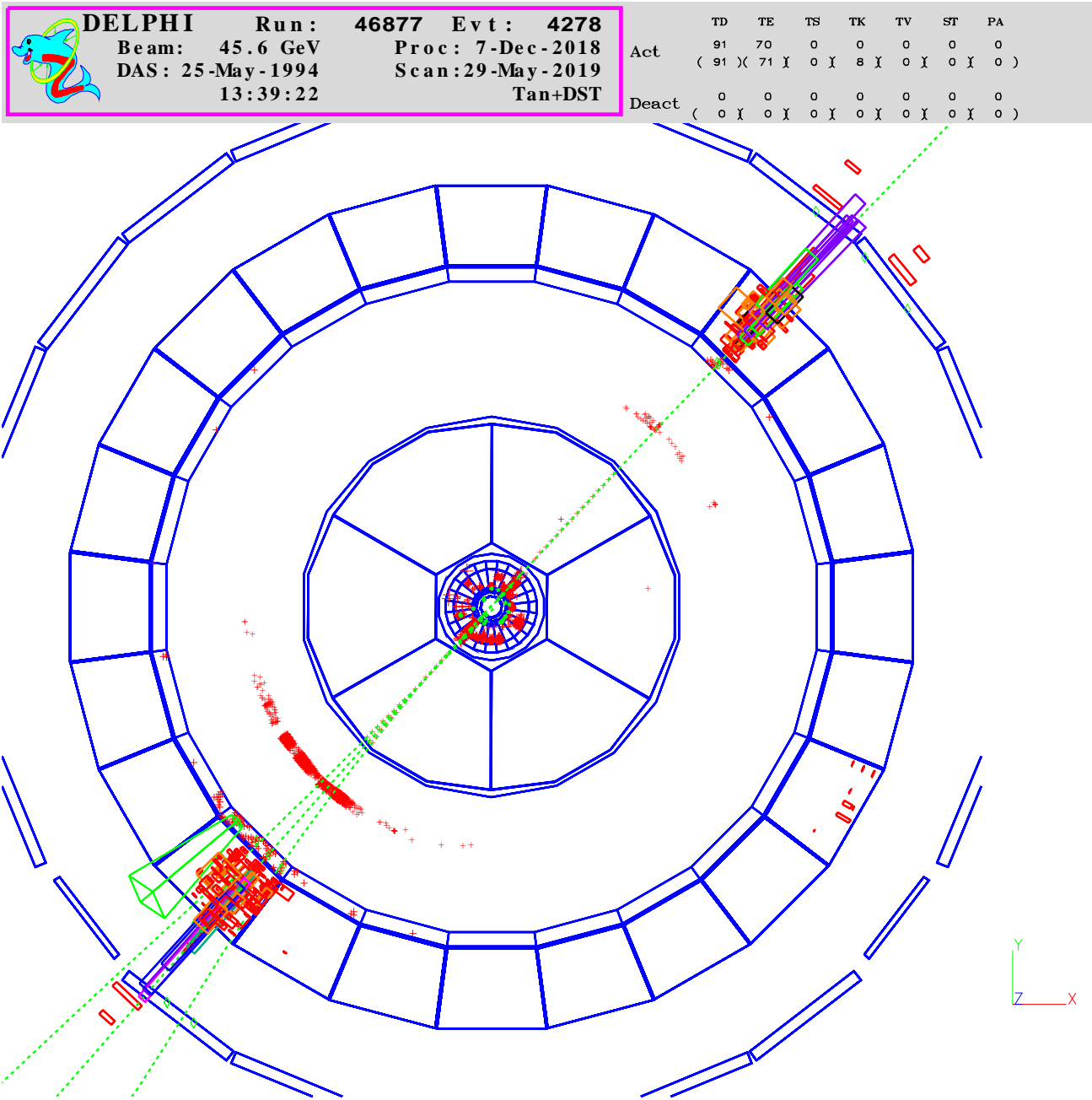,bbllx=50pt,bblly=250pt,bburx=500pt,bbury=600pt,%
width=14cm,angle=0}
\end{center}
\vskip5cm
\caption{ An event of topology 3, a general view.}
\label{fig:63}
\end{figure}

\newpage
\begin{figure}
\begin{center}
\epsfig{file=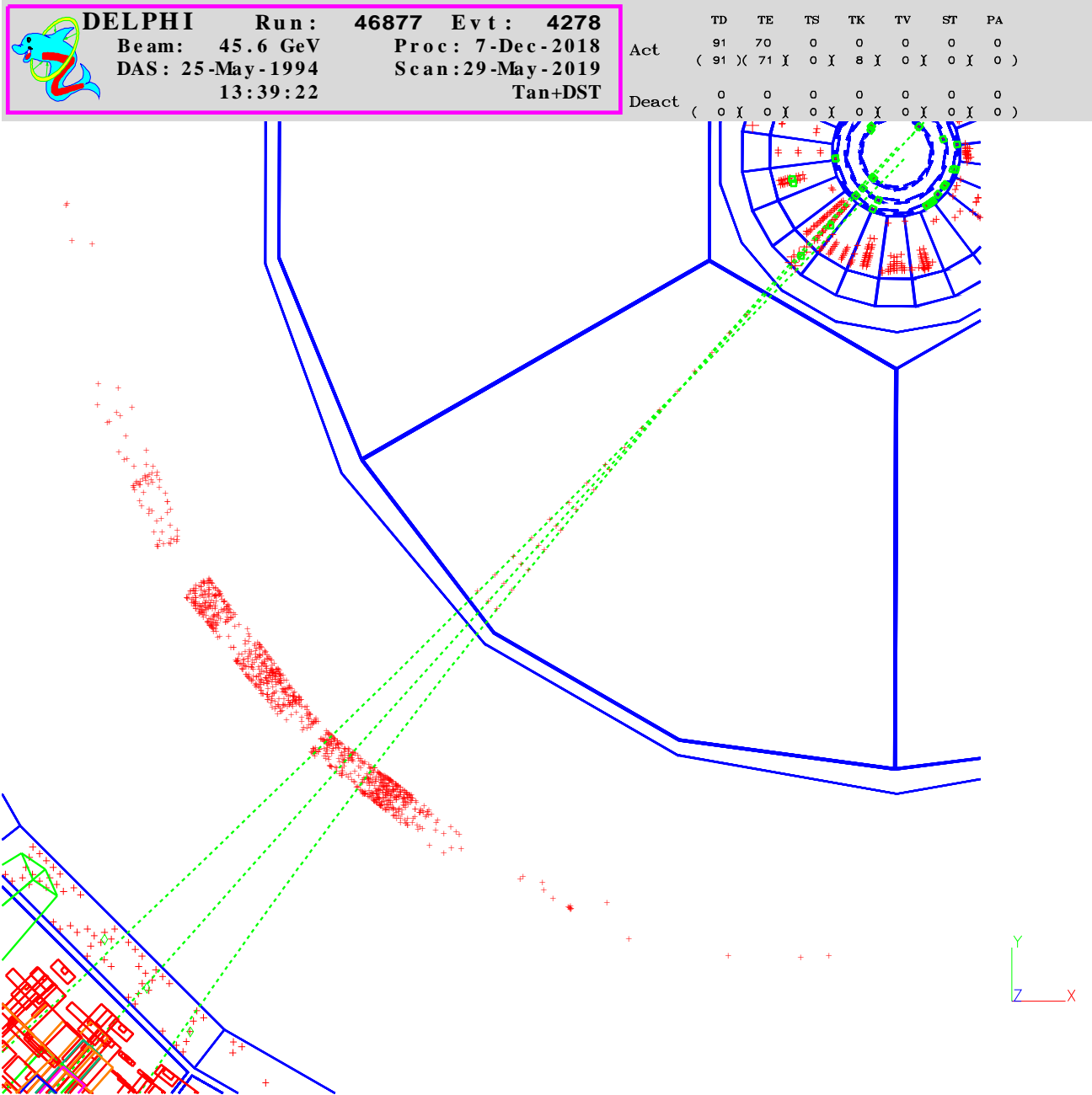,bbllx=50pt,bblly=250pt,bburx=500pt,bbury=600pt,%
width=14cm,angle=0}
\end{center}
\vskip5cm
\caption{ The same event, the 3-particle jet view.}
\label{fig:163}
\end{figure}

\newpage
\begin{figure}
\begin{center}
\epsfig{file=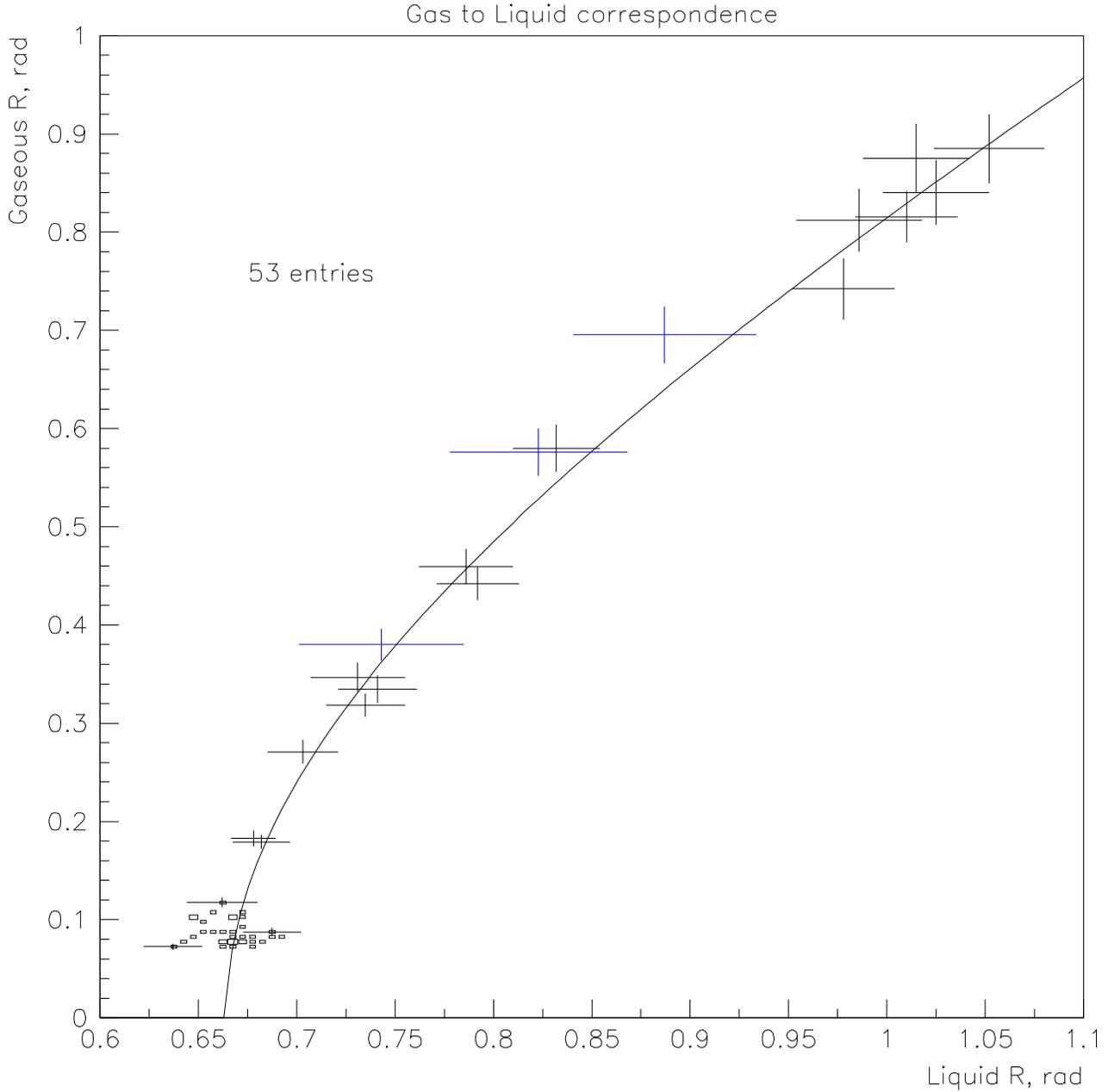,bbllx=50pt,bblly=180pt,bburx=550pt,bbury=670pt,%
width=17cm,angle=0}
\end{center}
\vskip0.3cm
\caption{ Radii of the anomalous rings expressed in angular units
as measured in the liquid and in the gaseous radiators. The curved line shows
the expected theoretical correspondence between the radii in the two radiators
given by the formula $n_{liq} \cos{\theta_{liq}} = n_{gas} \cos{\theta_{gas}}$,
with $n_{liq} = 1.273$ and $n_{gas} = 1.00194$. The errors in the left bottom
corner are suppressed to avoid pile-up; only several crosses are left in order
to present a typical angular error in this region; 5 boxes of the area twice 
as big as that of the rest boxes in this corner represent double entries.
3 crosses drawn in dark blue correspond to anomalous rings coming from the
quartz radiator. Their radii are recalculated to the liquid ring radii taking
into account the difference of the corresponding refraction indices.
For presentation purposes the scale along the $x$ axis is made two times finer 
than that of the $y$ axis.}
\label{fig:15}
\end{figure}

%\newpage
%\begin{figure}
%\begin{center}
%\epsfig{file=gasliqthtsml.ps,bbllx=50pt,bblly=180pt,bburx=550pt,bbury=670pt,%
%width=17cm,angle=0}
%\end{center}
%\vskip1cm
%\caption{ Radii of the anomalous rings as measured in the liquid and in the
%gaseous radiators from the left bottom corner of Fig.~\ref{fig:15}. 
%The expected theoretical correspondence between the radii
%in the both radiators is shown by a curve. 
%For presentation purposes the scale along the $y$ axis is made two times finer 
%than that of the $x$ axis.}
%\label{fig:15a}
%\end{figure}

\newpage
\begin{figure}
\begin{center}
\epsfig{file=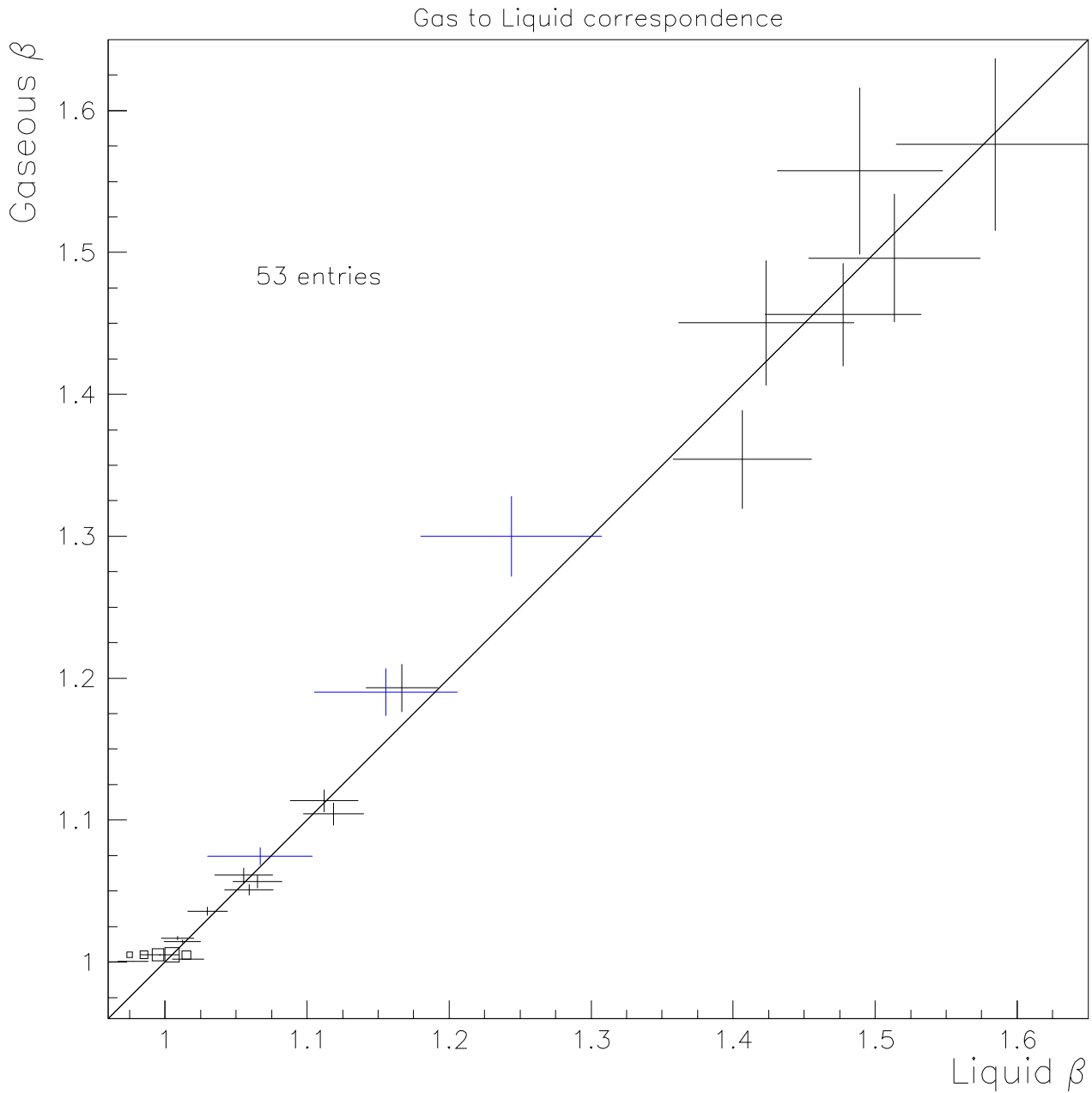,bbllx=50pt,bblly=180pt,bburx=550pt,bbury=670pt,%
width=17cm,angle=0}
\end{center}
\vskip2cm
\caption{Particle velocities ($\beta = v/c$) corresponding to radii of 
the anomalous rings as measured in the liquid and in the gaseous radiators. 
The diagonal indicates where the values of the two variables 
%$1/(n_{liq}\cos{\theta_{liq}})$ and  $1/(n_{gas}\cos{\theta_{gas}})$, with
%$n_{liq} = 1.273$ and $n_{gas} = 1.00194$, 
$\beta_{liq}$ and $\beta_{gas}$ are equal. 
3 crosses drawn in dark blue correspond to anomalous rings coming from 
the quartz radiator (see caption to Fig.~\ref{fig:15}).
The correlation coefficient for the points in this plot is 0.992. The sum of 
$\chi^2$ for the deviations of the 53~points from the diagonal is 40.1. 
The corresponding p-value for the points to be consistent with the diagonal 
is near~95\%.}
\label{fig:16}
\end{figure}

\newpage
\begin{figure}
\begin{center}
\epsfig{file=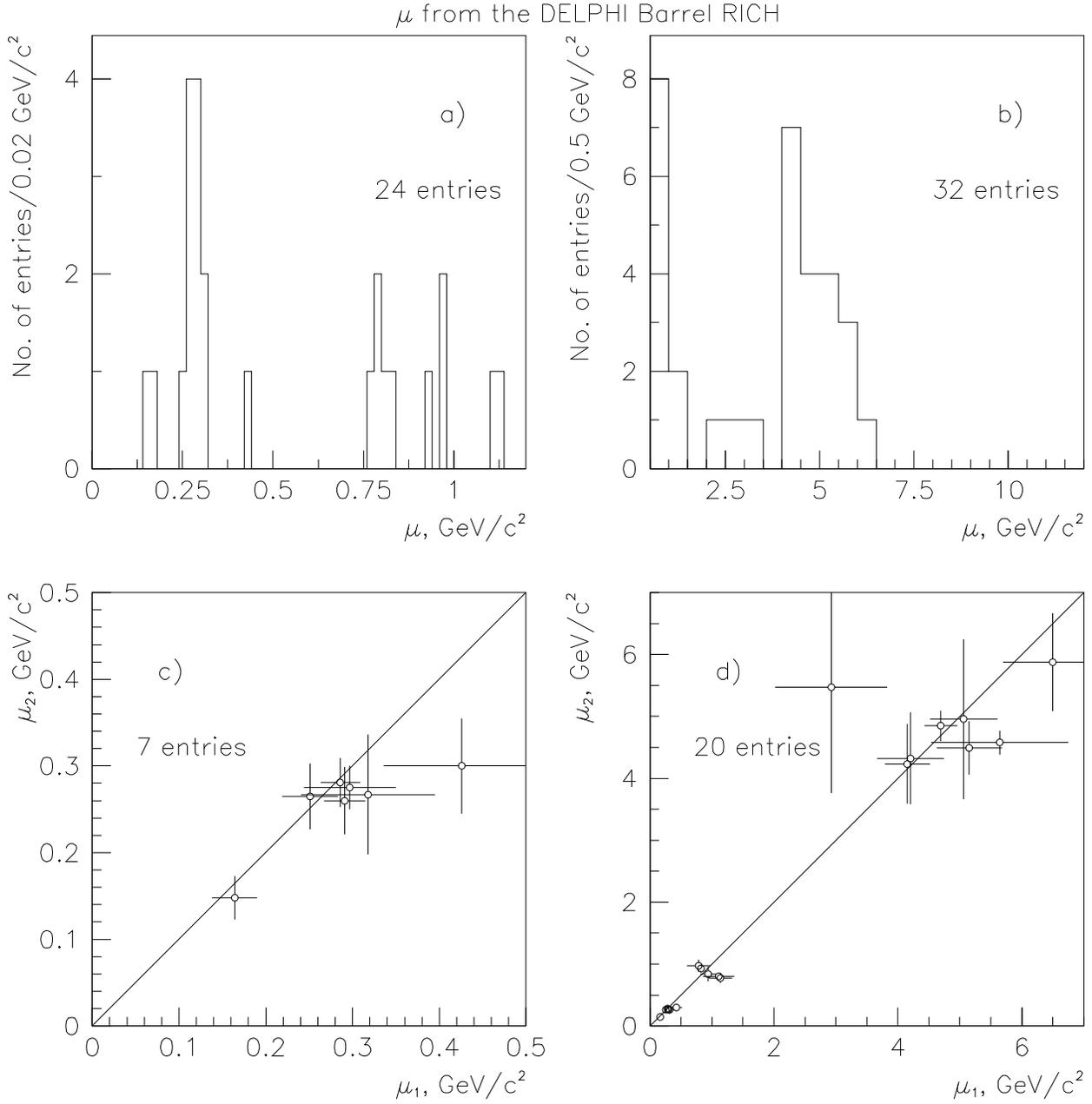,bbllx=50pt,bblly=180pt,bburx=550pt,bbury=670pt,%
width=17cm,angle=0}
\end{center}
\vskip1cm
\caption{{\bf a)} The distribution of the tachyon mass parameters as calculated
from the measured Cherenkov angles and the track momenta, low mass parameter
region; {\bf b)} The same for a high mass parameter region;
{\bf c)} Two-dimensional plot of tachyon mass parameters for events in which
both tachyon candidate track momenta were measured, low mass parameter region;
{\bf d)} The same for the overall mass parameter region. The diagonals in 
panels {\bf c)} and {\bf d)} indicate where the values of the two variables
$\mu_1$ and $\mu_2$ are equal.}
\label{fig:17}
\end{figure}
\newpage
\begin{figure}
\begin{center}
\epsfig{file=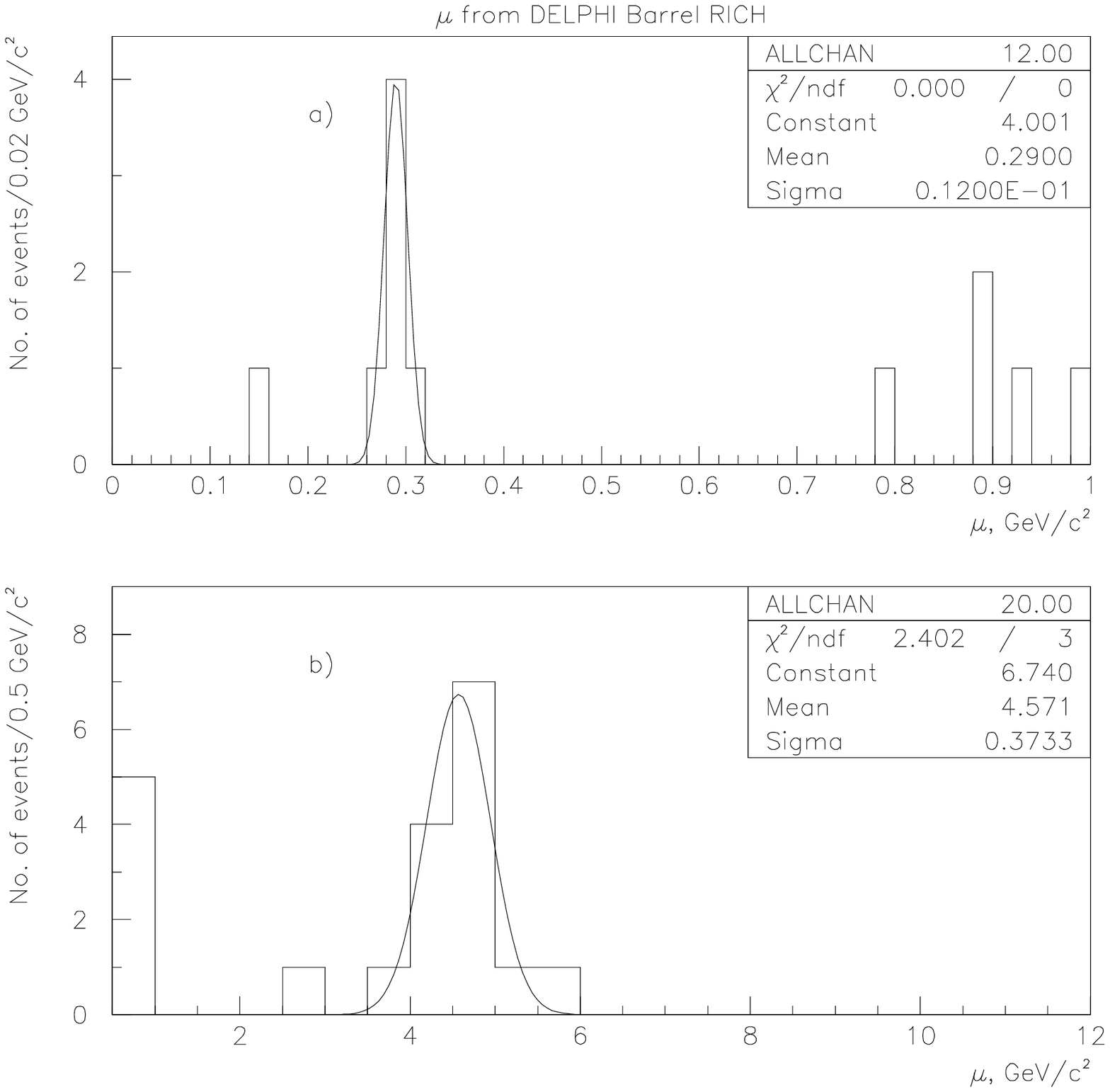,bbllx=50pt,bblly=180pt,bburx=550pt,bbury=670pt,%
width=17cm,angle=0}
\end{center}
%\vskip1cm
\caption{The distribution of the tachyon mass parameters obtained with
events containing anomalous rings and passed kinematic fits: {\bf a)} in the
mass region below 1~GeV/$c^2$ (to be compared with Fig.~\ref{fig:17}a). 
The curve presents the result of a fit of the peak at 0.29~GeV/$c^2$ by 
a Gaussian; {\bf b)} an analogous tachyon mass parameter distribution in 
the high mass region, the low mass region (below 0.5~GeV/$c^2$) being excluded 
(to be compared with Fig.~\ref{fig:17}b). The curve presents the result 
of a fit of the peak at 4.5 GeV/$c^2$ by a Gaussian.}
\label{fig:18}
\end{figure}

\newpage

\newpage
\begin{thebibliography}{ References}
%\vskip-3mm
\bibitem{prashort} V. F. Perepelitsa, T. Ekelof, A. Ferrer, B. R. French,
{\em A search for anomalous Cherenkov rings},
arXiv:hep-ex/1912.11839 (2019)
\bibitem{bds} O. M .P. Bilaniuk, V. K. Deshpande, E. C. G. Sudarshan,
{\em ``Meta"-relativity}, 
Amer. J. Phys. {\bf 30} (1962) 718-723.
\bibitem{fein} G. Feinberg,
{\em Possibility of faster-than-light particles},
Phys.Rev. {\bf 159} (1967) 1089-1105.
\bibitem{wigner2} E. P. Wigner, {\em Invariant quantum mechanical equations 
of motion}, in a book {\em Theoretical Physics}, pp.~59-82, I.A.E.A., Vienna,
(1963)
\bibitem{pdg} Particle Data Group, Tanabashi et al., Phys. Rew.~D {\bf 98}, 
No.~3, p.~414 (2018).
\bibitem{sigal} R. Sigal, A. Shamaly, 
{\em Tachyon behavior in general relativity}, 
Phys. Rev. D {\bf 10}, 2358-2361 (1974).
\bibitem{pvmich} V. P. Perepelitsa, {\em Tachyon Michelson experiment},
Phys. Lett. B {\bf 67}, 471-473 (1977).
\bibitem{pvcaus} V. F. Perepelitsa, {\em Causality, Relativity and
Faster-Than-Light Signals}, in a book {\em Philosophical Problems of
the Hypothesis of Superluminal Velocities} (Nauka Press, Moscow, 1986), p. 40
\bibitem{rembl} J. Rembieli\'{n}ski, 
{\em Tachyons and preferred frames},
Int. J. Mod. Phys. A {\bf 12} (1997) 1677-1710.
\bibitem{radzik} M. J. Radzikowski,
{\em Stable, renormalizable, scalar tachyonic quantum field theory 
with chronology protection}, 
arXiv:math-ph/0804.4534 (2008).
\bibitem{ttheor} V. F. Perepelitsa,
{\em Looking for a theory of faster-than-light particles}, 
arXiv:gen-ph/1407.3245 (2014).
\bibitem{causal} V. F. Perepelitsa, 
{\em How to implant a causal $Theta$-function into a tachyon field operator, 
or why tachyons do not violate causality }, 
arXiv:gen-ph/1512.07921 (2015).
\bibitem{delphi1} DELPHI Collaboration, P.~Aarnio et al., Nucl. Instr. and Meth. A {\bf 303} (1991) 233.
\bibitem{delphi2} DELPHI Collaboration, P.~Abreu et al., Nucl. Instr. and Meth. A {\bf 378} (1996) 57.
\bibitem{rich1} W. Adam et al., Nucl. Instr. and Meth. A {\bf 343} (1994) 68.
\bibitem{rich2} W. Adam et al., Nucl. Instr. and Meth. A {\bf 360} (1995) 416.
\bibitem{rich3} W. Adam et al., Nucl. Instr. and Meth. A {\bf 367} (1995) 233.
\bibitem{rich4} W. Adam et al., Nucl. Instr. and Meth. A {\bf 371} (1996) 12.
\bibitem{rich5} E. Albrecht et al., Nucl. Instr. and Meth. A {\bf 433} (1999) 
47.
%\bibitem{bloch} D. Bloch, M. Dracos, S. Tsamarios, Nucl. Instr. and Meth. A
%{\bf 371}, 236 (1996)
%\bibitem{perez} J. Garc\'{i}a P\'{e}rez, A. L\'{o}pez Ag\"uera, Nucl. Instr.
%and Meth. A {\bf 371}, 232 (1996); ibid, {\bf 343}, 276 (1994)
%\bibitem{delana} DELANA User's guide, DELPHI note, DELPHI 89-44 PROG 137 (1989)
\bibitem{kluit} M. Battaglia, P. M. Kluit, DELPHI note, DELPHI 96-133 RICH 90, 
http://inspirehep.net/record/1659476
\bibitem{thexp}  V. F. Perepelitsa,{\em Experimental aspects of the tachyon
hypothesis}, arXiv:gr-qc/1606.04808 (2016)
%\bibitem{unita2} Jacobson T., Tsamis N.~C., Woodard R.~P., {\em Tachyons and
%perturbative unitarity}, Phys. Rev. D {\bf 38}, 1823-1834 (1988)
%\bibitem{vip} A. Di Meglio et al., CERN Openlab White Paper ``Future ICT
%Challenges in Scientific Research" (2017)~29. 
%\bibitem{compton} S. Paiano, A. Perrotta, DELPHI note, DELPHI 98-37 CAL 139,
%http://inspirehep.net/record/1661282
%\bibitem{test2} V.f. Perepelitsa, T. Ekelof, A. Ferrer, B.R. French,
%{\em Study of Cherenkov rings in Bhabha and $\gamma - \gamma$ events with the 
%DELPHI Barrel RICH}, in preparation.
%\bibitem{ncontrol} et al.,
%\bibitem{glep1} DELPHI Collaboration, P.~Abreu et al., Phys. Lett. B {\bf 327}
%(1994) 386.
%\bibitem{glep2} DELPHI Collaboration, J.~Abdallah et al., Euro. Phys.~J. C 
%{\bf 37} (2004) 405.
%\bibitem{elep1} DELPHI Collaboration, P.~Abreu et al., Euro. Phys.~J. C
%{\bf 16} (2000) 371.
%\bibitem{elep2} DELPHI Collaboration, J.~Abdallah et al., Euro. Phys.~J. C 
%{\bf 45} (2006) 589.
\bibitem{alice} ALICE Collaboration, B. Abelev et al., Int. J. Mod. Phys A 
{\bf 29}, 1430044 (2014)
\bibitem{lhcbdet} LHCb Collaboration, A. A. Alves Jr. et al., J. Instrum.
{\bf3}, (2008) S08005.
\bibitem{lhcb} LHCb RICH Collaboration, M. Adinoli et al., Eur. Phys. J. C
{\bf 73} (2015) 2431.
\end{thebibliography}
\end{document}